\documentclass[preprint,3p,times]{elsarticle}

\usepackage{lineno,hyperref}
\modulolinenumbers[5]

\usepackage{xcolor}
\usepackage{subcaption}
\usepackage{amsmath}
\usepackage{multicol}
\usepackage{wrapfig}
\RequirePackage{tabularx}
\usepackage{booktabs} 
\usepackage{xcolor,colortbl}
\usepackage{ colorspace}
\usepackage[toc,page]{appendix}
\usepackage{caption}
\usepackage{multirow}
\definecolor{Gray}{gray}{0.85}
\definecolor{beaublue}{rgb}{0.74, 0.83, 0.9}
\definecolor{columbiablue}{rgb}{0.61, 0.87, 1.0}
\definecolor{lightbluetrial}{rgb}{0.68, 0.85, 0.9}
\colorlet{lightblue}{lightbluetrial!30}
\definecolor{non-photoblue}{rgb}{0.64, 0.87, 0.93}

\usepackage{mathtools}
\newlength{\bracewidth}

\newcolumntype{a}{>{\columncolor{lightblue}}c}
\newcolumntype{b}{>{\columncolor{white}}c}
\newcolumntype{k}{>{\columncolor{white}}c}

\newcommand{\simulink}{\textsc{SIMULINK}\textsuperscript{\textregistered} }
\newcommand{\matlab}{\textsc{MATLAB}\textsuperscript{\textregistered} }
\newcommand{\msc}{\textsc{MSC}\textsuperscript{\textregistered} }

\begin{document}

\begin{frontmatter}

\title{Sub-Structuring Modeling of Large Space Truss Structures for \\ Structure/Control Optimization in Presence of Parametric Uncertainties}

\author{A. Finozzi\fnref{ISAE}}
\ead{antonio.finozzi@student.isae-supaero.fr}
\author{F. Sanfedino \corref{mycorrespondingauthor}\fnref{ISAE}}
\ead{francesco.sanfedino@isae.fr}
\author{D. Alazard \fnref{ISAE}}
\ead{daniel.alazard@isae.fr}
\address{10 Avenue Edouard Belin, BP-54032, 31055, Toulouse, Cedex 4, France}
\fntext[ISAE]{Institut Sup\'erieur de l’A\'eronautique
	et de l’Espace (ISAE-SUPAERO),
	Universit\'e de Toulouse}


\cortext[mycorrespondingauthor]{Corresponding author}


\begin{abstract}
Modern and future high precision pointing space missions face increasingly high challenges related to the widespread use of large flexible structures. The development of new modeling tools which are able to account for the multidisciplinary nature of this problem becomes extremely relevant in order to meet both structure and control performance criteria. This paper proposes a novel methodology to analytically model large truss structures in  a sub-structuring framework. A three dimensional unit cube element has been designed and validated with a Finite Element commercial software. This model is composed by multiple two-dimensional sub-mechanisms assembled using block-diagram models. This constitutes the building block for constructing complex truss structures by repetitions of the element. The accurate vibration description of the system and its minimal representation, as well as the possibility of accounting for parametric uncertainties in its mechanical parameters, make it an appropriate tool to perform robust Structure/Control co-design. In order to demonstrate the strengths of the proposed approach, a co-design study case is proposed by combining a multidisciplinary optimization approach based on particle swarm algorithm and multiple structured robust $H_\infty$-synthesis. This has been used to optimize the pointing performances of an high pointing antenna, minimizing the perturbations coming from the Solar Array Mechanisms (SADM) of two solar panels, performing active control by means of multiple proof mass actuators, and simultaneously reduce the mass of the truss-structure which connects the antenna to the main spacecraft body. 

\end{abstract}


\end{frontmatter}

\section{Introduction}



In order to systematically face the challenges associated with the next generation of satellites, the European Space Agency (ESA) and NASA have combined their past experiences to cope with the fine pointing requirements of high accuracy observation and Science missions \cite{dennehy2018spacecraft}. This represent a domain which is extremely multi-disciplinary:  structural, control and system engineering considerations must coalesce to limit the propagation and amplification of internally generated disturbances through the satellite's flexible structures. For these reasons, the development of rigorous methodologies and design tools that can handle all these domains is crucial at early stages of design.

In the past decades, structural and control co-design has attracted a lot of attention due to its ability of merging these multiple multidisciplinary requirements into a single design flow. Moreover, the increasing use of large structures and appendages for Space applications has rendered flexible modal analysis mandatory for the design of proper spacecraft control laws.

In order to tackle the non-trivial modeling and analysis of these large and complex space systems, a sub-structuring technique using a multi-body approach is often considered to conceptually simplify the model. Seeing the overall structure as an assembly of multiple simpler sub-systems with increasing complexity has also the advantages of handling different types of boundary conditions at block assemblage level and easy sub-system validation.

The wide use of this approach for space applications has raised a significant interest in the development of proper modeling techniques that can prove to be versatile enough to account for multiple multi-body configurations, ranging from open-loop chains to closed-loops mechanisms.

Many sub-structuring techniques can be found in literature. A common approach relies on approximations linked to the Finite Element Method (FEM) or the Assumed Modes Method (AMM) \cite{Theodore1995}. However, these methods are heavily influenced by the set of predetermined boundary conditions assigned to the model, which may be drastically variable, for example in mass time-varying systems. Nevertheless, the FEM is currently the most popular method in structural analysis and it has had a deep impact in almost all multi-body modeling methods, such as the transfer matrix (TM) method and the component modes synthesis (CMS).

The Transfer Matrix (TM) Method, introduced by Holzer \cite{holzer1921analysis} and later independently by Myklestad \cite{myklestad1945new} creates a transfer matrix that links up the state vectors (generalized accelerations and forces) of the two extremities of the flexible body. These methods are particularly well suited for serially connected bodies and open-chain structures. Their major drawback is the inversion problems of the matrices composing the model \cite{tan1990modified}, whose matrices may be non-square or non-invertible depending on the boundary conditions. Moreover, these approaches are not optimal for a multi-body tree-like structures, where multiple inputs and multiple outputs are required at each end of the model \cite{Rui2014}. This drawback is particularly significant for space applications, where multiple bodies and flexible appendages, such as antennas and solar panels, are attached to a central spacecraft body. 

Component mode synthesis (CMS) approaches have been quite popular in the modeling of multi-body systems \cite{hurty1965dynamic,macneal1971hybrid,hintz1975analytical} thanks to  matrix condensation reduction, which renders this method particularly suited for sub-structuring problems \cite{holm2009component, tran2014reduced, yu2016element}. 
However, in this approach, the sub-component matrices overlap with each other to create a complex model, difficult to understand and to extend to systems with varying sizing parameters.

Methods based on effective mass/inertia of the appendages \cite{Girard2008} and effective impedance matrix \cite{pascal1987dynamics} represent another viable option to represent multi-body systems
and have had significant space applications in \cite{cumer2014codesign, guy2014dynamic,tantawi2008linear} for attitude control purposes.
These methods, however, lose the complete vibration behavior description of the systems, as they aim at delivering only the dynamic relation of state variables at the appendage root point. The application of these methods to design chain-like mechanisms is therefore not possible. 


The Two-Input-Two-Output Port (TITOP) Model, a direct dynamic approach initially proposed in \cite{alazard2015two}, overcomes these issues. The structure is conceived as a minimal state-space transfer between the accelerations and wrenches at the extremity points of the appendage and embeds both the direct and inverse dynamics: the IN/OUT channels are easily numerically invertible to account for multiple boundary conditions. Moreover this approach in a block-diagram model permits the design of closed-chain multi-body systems for any boundary conditions by creating feedback loops and inverting IN/OUT channels. An analytical uniform beam model in the TITOP approach was proposed by Mural \textit{et al.} \cite{Murali2015}, while Perez \textit{et al.} \cite{perez2015linear, perez2015flexible} introduced model parametrization in the Linear Fractional Transformation (LFT) form for co-design and robust control applications. A complete formulation of the TITOP beam model was proposed by Chebbi \textit{et al.} \cite{Chebbi2016}, who performed a rigorous comparison with the Euler-Bernoulli beam theory for all boundary conditions. This work provided the first application of the TITOP model with inverted channels, which was used to model the four-bar mechanism by assembling multiple TITOP blocks. Sanfedino \textit{et al.} \cite{SANFEDINO2018} validated the inversion operation not only for analytical TITOP models but also for numerical ones provided by FEM commercial software and then extended the TITOP formalism for N-Input-N-Output Port (NINOP) models. 


All the models derived in TITOP approach have been implemented in the last release of the Satellite Dynamics Toolbox (SDT) \cite{SDT1,d_alazard_f_sanfedino_satellite_2021}, which allows the user to easily build  models of multi-body systems, fully compliant with the \matlab routines of the Robust Control Toolbox \cite{balas2007robust} to perform parametric robust control design and analysis.

All previous work, however, dealt only with simple systems when considering closed-loop kinematic chains. The main contribution of this work is to present a novel methodology to analytically model complex truss structures in the TITOP sub-structuring approach. These 3D system models represent a new set of tools which can be used to perform Structure/Control co-design, as they accurately describe the vibration dynamics response of the flexible bodies and have a minimal representation, limiting the number of DOFs present in the model.  In particular, following a procedure similar to the one found in \cite{Chebbi2016}, multiple TITOP elements are assembled to form complex mechanical systems that can later be re-assembled to form large scale structures: multiple two dimensional (2D) and three dimensional (3D) multi-body systems will be presented with the aim of assembling a cubic structural element. The latter represents the minimal model of a complex 3D structure which can be used as the building block for the construction of large space truss structures. 

In order to showcase the strength of the proposed approach, this work proposes a case study to display the ability of these parametric models of performing Structure/Control co-design. The dual optimization of the structural design and the control performance fits the framework of the Multidisciplinary Design Optimization (MDO), which originated from the work of Schmit and Haftka \cite{Schmit1960,Haftka1973}. The multidisciplinary approach is used when conflicting optimizations objectives stem from different disciplines and when a classical sequential optimization may fail to find the global optima of the problem. This is the case for Structure/Control co-design problems, where usually structural optimization is performed before the control one, iterating the process multiple times. This can be a long process and convergence is not granted due to the concurrent nature of the two sub-problems: mass reduction can increase significantly the flexibility of the system, whose low frequency modes may interfere with the satellite's Attitude Control System (ACS), as seen in \cite{falcoz2013integrated}.

Martins and Lambe \cite{martins2013multidisciplinary} presented the latest and most complete literature review on MDO and identified two main architectures in this framework: monolithic and distributed.

Monolithic architectures solve the MDO problem using a unique optimization process. They are proven to be more  efficient \cite{fathy2001coupling, reyer2001comparison} than classical optimization techniques \cite{chen20063d, li2001design}, especially when \textit{bidirectional coupling} is present between the structural and control sub-problem. This is the case when one problem depends on some variables or parameters of the other sub-problem \cite{frischknecht2011pareto}. 
Examples of this monolithic approach in this field can be found in \cite{hale1985optimal, allison2014co, maraniello2016optimal}. 

Distributed architectures, on the other hand, are more adapted for optimization problems where \textit{unidirectional coupling} takes place. For this family of problems, the structural mechanical sub-problem depends solely on the mechanical design parameters of the system, while the control one depends both on them and the control design parameters, as seen in \cite{frischknecht2011pareto}. Nested optimization techniques can be used to solve this type of optimization problems, as showcased in the works of Chilian \textit{et al.} \cite{chilan2017co} and by the BIOMASS test case \cite{falcoz2013integrated, toglia2013optimal}.  Multiple optimization algorithms are found in literature to solve the nested optimization problem: gradient-based methods, as in \cite{kraft1988software}, and global optimization methods (particle swarm \cite{kennedy1995particle}, genetic algorithm \cite{deb2011multi} and pattern search \cite{audet2002analysis}).

When an analytical model of the structure is available, the integration of the design variables as uncertain parameters opens the possibility of achieving control/structure co-design in a unique iteration, using the non-smooth techniques available in the robust control framework \cite{gahinet2011structured},  as shown in \cite{alazard2013avionics,oatao18685,perez2015linear} on Aerospace applications. This monolithic direct co-design, however, does not result in the global optimal design, risking to fall within one of the many local minimums. 

This work proposes a novel nested optimization approach to solve the structure/control co-design problem presented in the case study. The problem can be identified as unidirectional, therefore a particle swarm optimization algorithm  can be used to perform the outer structural optimization, while the nested optimizations related to the control systems are handled within the structured $H_\infty$ framework.

The main contibutions of this paper are:
\begin{itemize}
	\item the development of a sub-structures model fully parameterized according to the mechanical sizing parameters and allowing the complex truss structure to be assembled by block-diagram interconnection,
	\item the validation of the 3D-cube element by a detailed comparison with the NASTRAN model,
	\item the co-design of a complex space truss structure holding an optical payload and of the LOS (Line Of Sight) stabilization controller.
\end{itemize}
After a brief introduction on the TITOP approach used to model a flexible appendage in section \ref{sec:TITOP}, the modeling of all 2D mechanism is detailed in section \ref{sec:2DMech}. The introduction and validation of the cubic structural element is performed in section \ref{sec:3DMech}. The in-depth description of the co-design study case is performed in section \ref{sec:Application_Case_Study}, while the optimal design is finally presented in section \ref{sec:results}.

\section{TITOP Approach}
\label{sec:TITOP}

\begin{figure}[htp!]
	\centering
	\begin{subfigure}{.64\textwidth}
		\includegraphics[width =.99\textwidth]{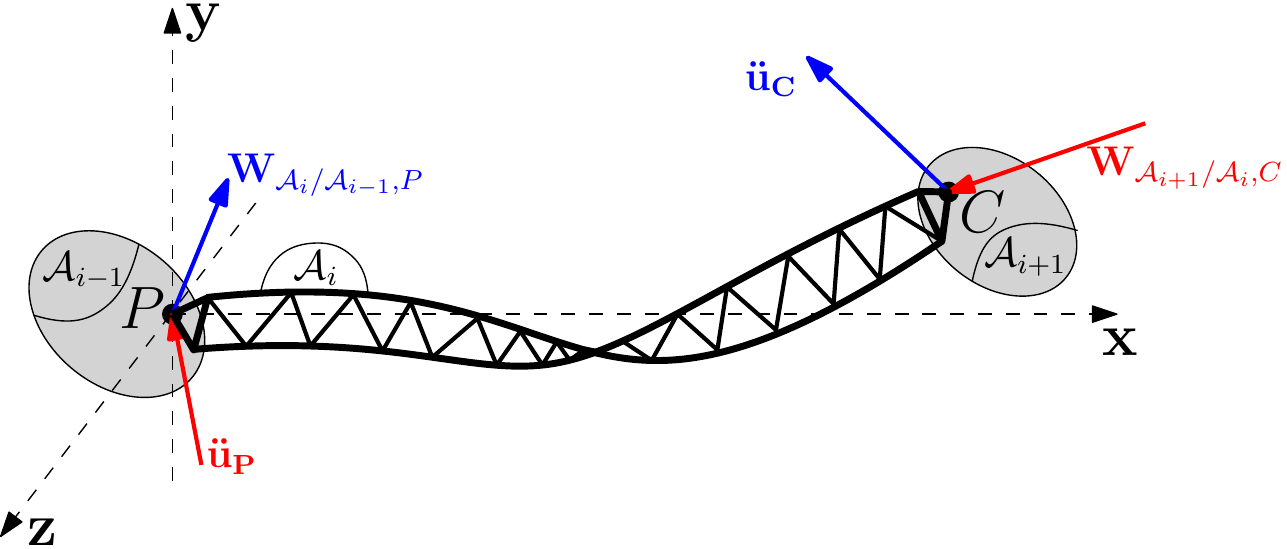}
	\end{subfigure}
	\begin{subfigure}{.35\textwidth}
		\includegraphics[width =1\textwidth]{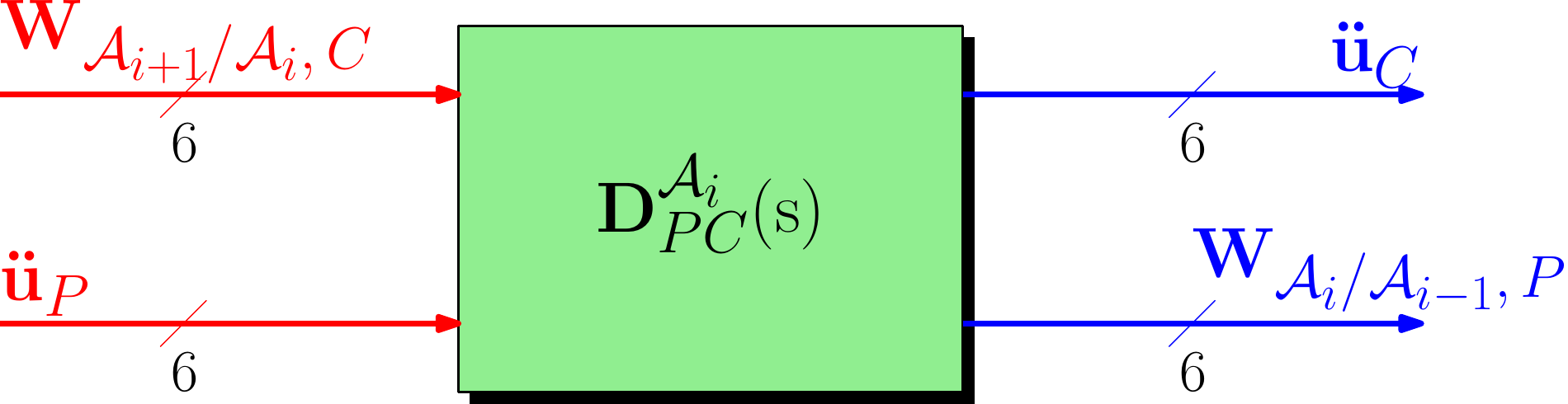}
	\end{subfigure}
	\caption{TITOP scheme and nomenclature for a generic flexible appendage $\mathcal{A}_i$.}
	\label{fig:TITOP_Beam}
\end{figure}

Let us consider a flexible body $\mathcal{A}_i$ as seen in Fig. \ref{fig:TITOP_Beam} (left) connected to a parent structure $\mathcal{A}_{i-1}$ at the point $P$ and to a child structure $\mathcal{A}_{i+1}$ at the point $C$. The resulting TITOP model $\mathbf{D}^{\mathcal{A}_i}_{PC}(\mathrm s)$, schematized in Fig. \ref{fig:TITOP_Beam} (right), is a $\{ 12\times12 \}$ linear dynamic model whose inputs are:
\begin{itemize}
	\item $\mathbf{W}_{\mathcal{A}_{i+1}/\mathcal{A}_{i},\, C} $: the $\{6\times 1 \}$ wrench (forces and torques) applied by the body $\mathcal{A}_{i+1}$ to $\mathcal{A}_{i}$ at point C;
	\item $\mathbf{\ddot{u}}_P$: the $\{6\times 1 \}$ inertial acceleration (linear and angular)  imposed by the parent body $\mathcal{A}_{i-1}$ at point $P$ to $\mathcal{A}_{i}$;
\end{itemize}
and the \textit{conjugated} outputs are:
\begin{itemize}
	\item $\mathbf{\ddot{u}}_C$: the $\{6\times 1 \}$ components of the inertial  acceleration of point $C$;
	\item $\mathbf{W}_{\mathcal{A}_{i}/\mathcal{A}_{i-1},\, P} $: the $\{6\times 1 \}$ wrench applied by $\mathcal{A}_{i}$ to the parent structure $\mathcal{A}_{i-1}$ at point P.
\end{itemize}
	All these input/output variables are projected in the body frame. Thus, to lighten the notations,  the projection frame is not mentioned in the various block-diagrams presented hereafter.

As described in \cite{Chebbi2016}, the state-space representation of  $\mathbf{D}^{\mathcal{A}_i}_{PC}(\mathrm s)$ can be directly built from the data (flexible mode frequencies, modal participation factors and modal shapes) of the \textit{clamped at $P$ - free at $C$} model of the body $\mathcal A_i$. Then channel inversion operations can be used to derive the model under different boundary conditions. 

Let us denote $[\mathbf M(\mathrm s)]^{-1_{\mathbf I}}$ the model $\mathbf M(\mathrm s)$ where the channel numbered in the vector of indexes $\mathbf I$ are inverted following the procedure described in Appendix 1 of \cite{Chebbi2016}, then for instance:
\begin{itemize}
	\item $[\mathbf{D}^{\mathcal{A}_i}_{PC}(\mathrm s)]^{-1_{\mathbf [1:6]}}$ models the body $\mathcal A_i$ under the \textit{clamped at $P$ - clamped at $C$} boundary conditions,
	\item $[\mathbf{D}^{\mathcal{A}_i}_{PC}(\mathrm s)]^{-1_{[1\;2\;3\;4\;6\,11]}}$ models the body $\mathcal A_i$ under the \textit{pinned at $P$  and $C$}  boundary conditions. Both pinned axes are the $ \mathbf y$-axis of the body frame used to project the model (Fig. \ref{fig:TITOP_Beam} (left)).
	\end{itemize}
A multi-body system composed of several flexible bodies and several revolute or clamped joints can then be built by the interconnection of the TITOP models of each body and the twice $6\times 6$ DCM (Direction Cosine Matrix) between the various body frames. This approach was embedded in the Satellite Dynamics Toolbox (SDT) with various features:
\begin{itemize}
	\item a NASTRAN/SDT interface to build the TITOP model directly from the ouput files (.f06 and .bdf) of the NASTRAN/PATRAN model,
	\item an analytical TITOP model of the \textsc{Euler-Bernoulli} beam in the 6 d.o.fs (degree-of-freedom)  case fully parametrized according to its length $l$ (along the $\mathbf x$-axis of the beam reference frame), its section area $S$, its second moments of area $I_z$ and $I_y$, its \textsc{Young} modulus $E$ and its mass density $\rho$. These parameters can be declared as varying parameters  to obtain an LPV (Linear Parameter Varying) model fully compliant  with the \matlab Robust Control Toolbox to perform  robust design and performance analysis.
\end{itemize}
This beam model, fully detailed in \cite{Chebbi2016}, is used as the basic element for the truss structure models presented in the sequel. The reader is advised to read the the SDTlib Users's Manual to have a deep insight in the SDT \cite{d_alazard_f_sanfedino_satellite_2021}.
\subsection{2D Mechanisms}
\label{sec:2DMech}
In this section, a series of two-dimensional mechanisms is presented. The main goal of these kinematics is to act as intermediate step towards building complex three-dimensional structures. By exploiting the assembly of multiple elementary TITOP beam blocks, several multi-body mechanical systems have been implemented: the so called \textit{L-Chain mechanism}, the \textit{Triangle mechanism} and the \textit{Square mechanism}. This section will detail the geometrical characteristics of these kinematics and their modeling using a block-diagram approach.

\subsubsection{L-Chain Mechanism}
\begin{figure}[htp!]
	\centering
	\begin{subfigure}{.32\textwidth}
		\includegraphics[width =\textwidth]{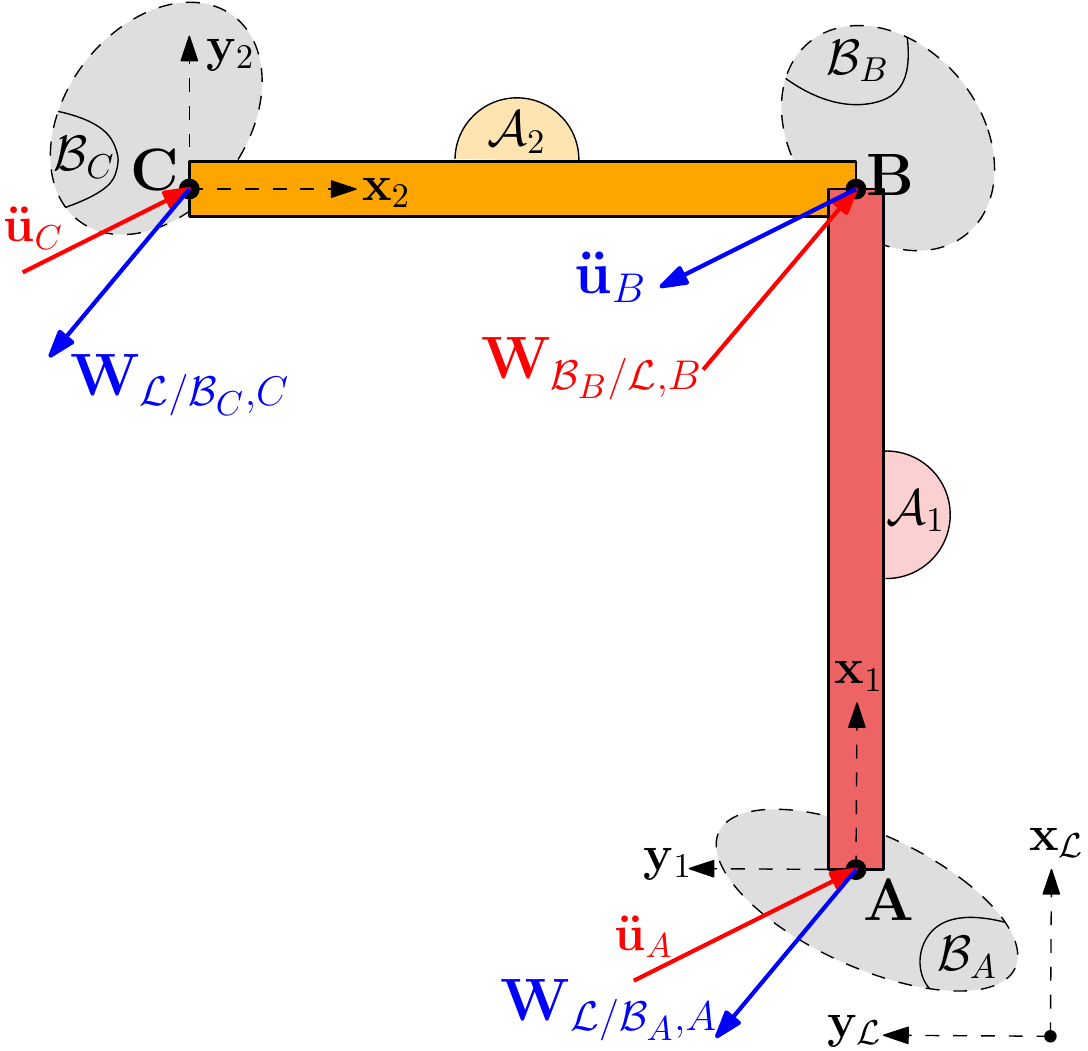}
		\caption{}
		\label{fig:L_Scheme}
	\end{subfigure}
	\begin{subfigure}{.67\textwidth}
		\includegraphics[width =\textwidth]{L_chain_blocks}
		\caption{}
		\label{fig:L_Sys}
	\end{subfigure}
	\caption{Representation of the \textit{Clamped-Free-Clamped} (CFC) L-Chain Mechanism $\mathcal{L}^{CFC}(s)$ (Fig. \ref{fig:Trig_Scheme}) and its block-diagram model (Fig. \ref{fig:Trig_Sys})}
	\label{fig:L-Mech}
\end{figure}

The L-Chain mechanism $\mathcal{L}$ is composed by two beams $AB$ (body $\mathcal{A}_1$) and $CB$ (body $\mathcal{A}_2$) connected at point $B$ in a given angular configuration $\alpha$ , as seen in Fig. \ref{fig:L_Scheme} (in the case $\alpha=\pi/2\,(rad)$). This mechanism is linked:
\begin{itemize}
	\item  to two parent bodies $\mathcal B_A$ and $\mathcal B_C$ at the point $A$ and $C$, imposing accelerations $\mathbf{\ddot{u}}_A$ and $\mathbf{\ddot{u}}_C$,
	\item to a child body $\mathcal B_B$  at point $B$ applying a wrench  $\mathbf{W}_{\mathcal{B}_{B}/\mathcal{L},\, B} $.
\end{itemize}
The $3$ input - $3$  output ports model of the mechanism $\mathcal L$ is then described by the block-diagram depicted in Figure  \ref{fig:L_Sys}. This model involves the \textit{clamped at $A$ - free at $B$} model $\mathbf{D}^{\mathcal{A}_1}_{A,\,B}(\mathrm s)$ of the beam $\mathcal{A}_1$ and the \textit{clamped at $C$ - clamped at $B$} model $[\mathbf{D}^{\mathcal{A}_2}_{A,\,B}(\mathrm s)]^{-1_{[1:6]}}$ of the beam $\mathcal{A}_2$. The upper ports of these to sub-model are connected in a feedback loop to take into account:
\begin{itemize}
	\item the loop closure constraint: the point $B$ on the 2  bodies must have the same accelerations $\mathbf{\ddot{u}}_B$ ,
	\item the wrench balance at the point $B$ of beam $\mathcal{A}_1$ : $ \mathbf{W}_{ ( \mathcal{A}_2 + \mathcal{B}_B )/ \mathcal{A}_1, B} = \mathbf{W}_{\mathcal{B}_B/\mathcal{L},B} - R_{2,1} \mathbf{W}_{\mathcal{A}_1/ \mathcal{A}_2, B}$.
\end{itemize}
Note that in Fig. \ref{fig:L-Mech} the blocks $R_{i,j}$ correspond to the twice $6\times 6$ DCM from the body frame of beam $\mathcal{A}_i$ to the body frame of beam $\mathcal{A}_j$.

This model is named  \textit{clamped-free-clamped (CFC) L-Chain} model and is denoted $\mathcal{L}^{CFC}(\mathrm{s})$. Indeed when the three inputs $\mathbf{\ddot{u}}_A$, $\mathbf{W}_{\mathcal{B}_{B}/\mathcal{T},\, B} $  and $\mathbf{\ddot{u}}_C$ (in this order)  are null, the $\mathcal{L}$ mechanism is clamped at point $A$, free at point $B$ and clamped at point $C$. This model described the dynamic behavior between three \textit{conjugated} input-output pairs associated to the three ports (connection points) of the mechanism. Thus, using the channel inversion operation, one can also defined:
\[
\mathcal{L}^{CCC}(\mathrm{s})=[\mathcal{L}^{CFC}(\mathrm{s})]^{-1_{[7:12]}},\quad \mathcal{L}^{FFC}(\mathrm{s})=[\mathcal{L}^{CFC}(\mathrm{s})]^{-1_{[1:6]}}, \cdots
\]
All the inputs and outputs of the model $\mathcal{L}^{CFC}(\mathrm{s})$ are projected in the body frame $(\mathbf x_\mathcal L,\; \mathbf y_\mathcal L,\; \mathbf z_\mathcal L)$ of the mechanism $\mathcal L$, chosen aligned with the frame of body $\mathcal A_1$ (see Fig. \ref{fig:L_Scheme}).

\subsubsection{Triangle Mechanism}
\begin{figure}[h!]
	\centering
	\begin{subfigure}{.37\textwidth}
\		\includegraphics[width =\textwidth]{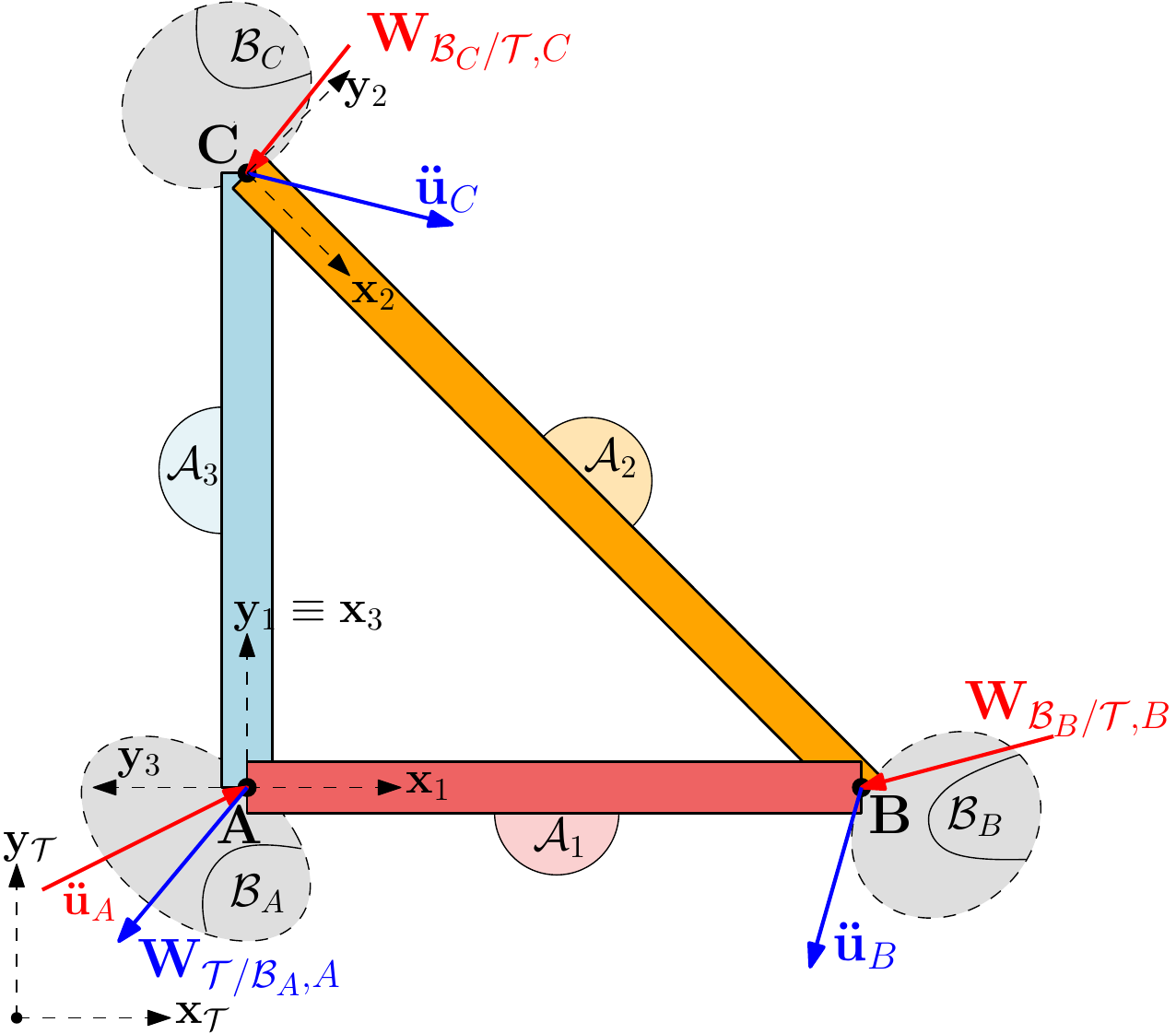}
		\caption{}
		\label{fig:Trig_Scheme}
	\end{subfigure}	
	\begin{subfigure}{.62\textwidth}
		\includegraphics[width =\textwidth]{Triangle_sys_block}
		\caption{}
		\label{fig:Trig_Sys}
	\end{subfigure}
	\caption{Representation of the \textit{Clamped-Free-Free} (CFF) Triangle Mechanism (Fig. \ref{fig:Trig_Scheme}) and corresponding block-diagram model of the system (Fig. \ref{fig:Trig_Sys})}
	\label{fig:TrigFig}
\end{figure}

The modeling of a basic triangular closed-loop mechanism is hereby considered. This mechanism is first modeled in the \textit{clamped at $A$, free at $B$, free at $C$} boundary conditions considering the acceleration  $\mathbf{\ddot u}_A$,  the wrenches $\mathbf{W}_{\mathcal{B}_{B}/\mathcal{T},\, B} $ and 
$\mathbf{W}_{\mathcal{B}_{C}/\mathcal{T},\, C} $ as inputs applied by the external bodies $\mathcal B_A$, $\mathcal B_B$ and $\mathcal B_C$ at points $A$, $B$ and $C$, respectively (see Fig. \ref{fig:Trig_Scheme}).

This model, denoted $\mathcal T^{CFF}(\mathrm s)$ can then be represented by the  block-diagram depicted in Fig. \ref{fig:Trig_Sys}. Indeed, since the loop closure constraint at the point $B$  is already taken into account in the model $\mathcal L^{CFC}(\mathrm s)$ previously presented (considering now that $\alpha=\pi/4\,(rad)$), the model $\mathcal T^{CFF}(\mathrm s)$  can be built by adding  the model $\mathbf{D}^{\mathcal{A}_3}_{A,\,C}(\mathrm s)$ of the third beam $AC$ (body $\mathcal A_3$)  and connecting it to the  model $\mathcal L^{CFC}(\mathrm s)$ to satisfy the new constraints:
\begin{itemize}
	\item the kinematic constraint: the point $A$ on the 2  bodies $\mathcal A_1$ and $\mathcal A_3$ must have the same accelerations $\mathbf{\ddot{u}}_A$ ,
	\item the wrench balance at the point $A$: $\mathbf{W}_{\mathcal{T}/\mathcal{B}_A, A} = \mathbf{W}_{{\mathcal{A}_1}/{\mathcal{B}_A}, A} + R_{3,\,1}\mathbf{W}_{{\mathcal{A}_3}/{\mathcal{B}_A}, A}$,
	\item the wrench balance at the point $C$: $	\mathbf{W}_{(\mathcal{B}_C + \mathcal{A}_2)/\mathcal{A}_3,C} = R_{1,\,3}\left(\mathbf{W}_{\mathcal{B}_C/\mathcal{T},C} +R_{2,\,1} \mathbf{W}_{\mathcal{A}_2/\mathcal{A}_3,C}\right)$.
\end{itemize}
All the inputs and outputs of this 3 input- 3 output port model $\mathcal{T}^{CFF}(\mathrm{s})$ are projected in the body frame $(\mathbf x_\mathcal T,\; \mathbf y_\mathcal T,\; \mathbf z_\mathcal T)$ of the mechanism $\mathcal T$, chosen aligned with the frame of body $\mathcal A_1$ (see Fig. \ref{fig:Trig_Scheme}).

\subsubsection{Square Mechanism}
\begin{figure}[htp!]
	\centering
	\begin{subfigure}{.34\textwidth}
		\includegraphics[width =\textwidth]{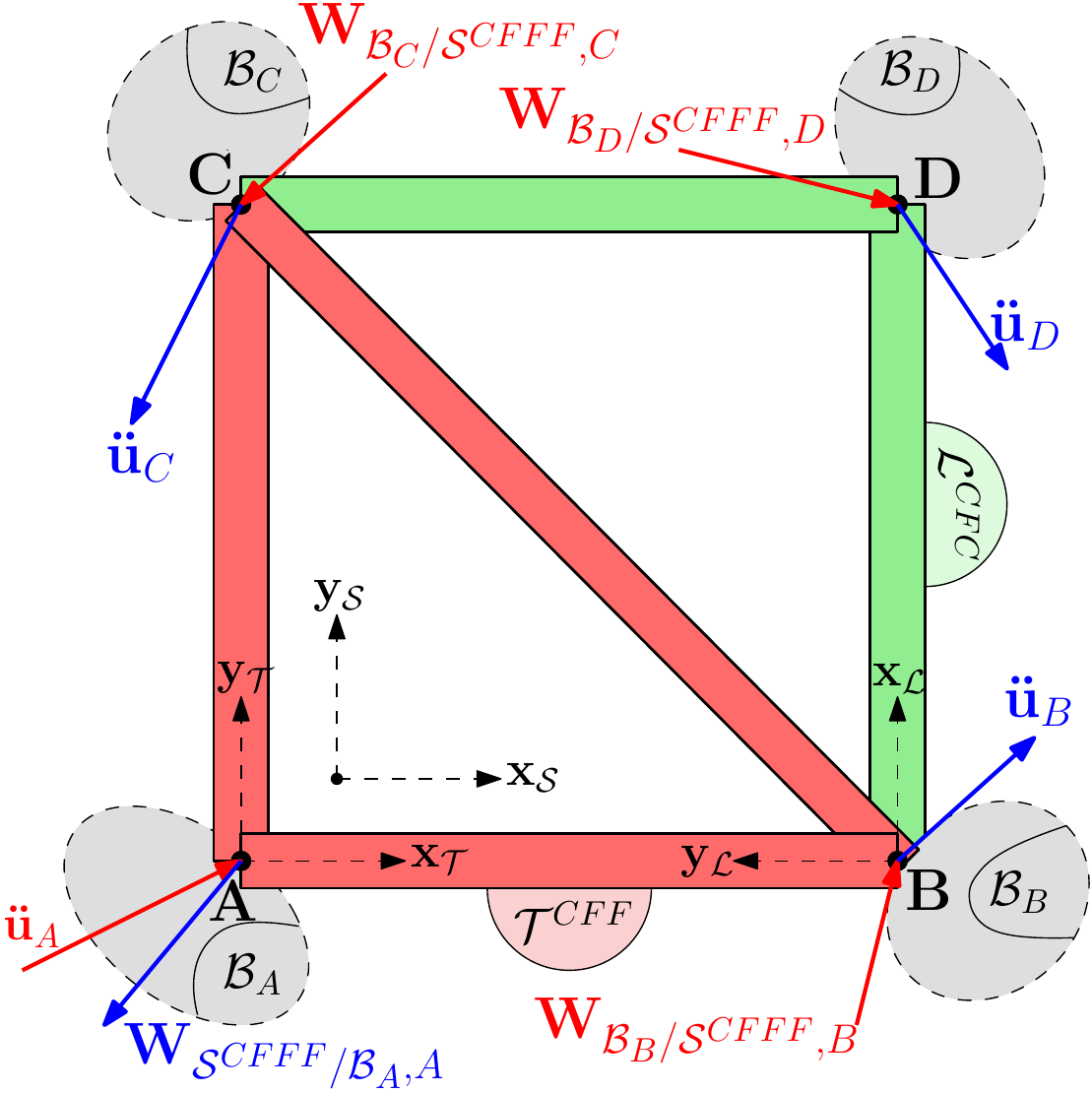}
		\caption{}
		\label{fig:Sq_Scheme}
	\end{subfigure}
	\begin{subfigure}{.65\textwidth}
		\includegraphics[width =\textwidth]{Square_sys_block.pdf}
		\caption{}
		\label{fig:Sq_Sys}
	\end{subfigure}
	\caption{Representation of the \textit{Clamped-Free-Free-Free} (CFFF) Square Mechanism $\mathcal{S}^{CFFF}(\mathrm{s})$ (a) and corresponding block-diagram model of the system (b)}
	\label{fig:Sq_Fig}
\end{figure}

The  square mechanism presented in Fig. \ref{fig:Sq_Fig} is  composed of five beams: four of them form a polygonal perimeter and the last one is positioned diagonally to create two closed loop chains. This mechanism is  modeled in the \textit{clamped at $A$, free at $B$, free at $C$,  free at $C$} boundary conditions considering the acceleration  $\mathbf{\ddot u}_A$,  the wrenches $\mathbf{W}_{\mathcal{B}_{B}/\mathcal{S},\, B} $, $\mathbf{W}_{\mathcal{B}_{C}/\mathcal{S},\, C} $ and 
$\mathbf{W}_{\mathcal{B}_{D}/\mathcal{S},\, D} $ as inputs applied by the external bodies $\mathcal B_A$, $\mathcal B_B$ , $\mathcal B_C$ and $ \mathcal B_D$ at points $A$, $B$ , $C$ and $D$, respectively (see Fig. \ref{fig:Sq_Scheme}). 

This model, denoted $\mathcal S^{CFFF}(\mathrm s)$ can then be represented by the  block-diagram depicted in Fig. \ref{fig:Sq_Sys}. It involves directly the  model $\mathcal T^{CFF}(\mathrm s)$ of the triangular mechanism and the model $\mathcal L^{CFC}(\mathrm s)$ of the L-chain mechanism with two feedback loops between their ports relative to the points $B$ and $C$, allowing to take into account kinematics constraints and wrench balances at these two connection points.  This allows for a drastic reduction in the  assembly complexity: the model is  composed only by two blocks while representing five flexible bodies in total. This showcases the power of this modular approach in structural design, whose advantages will be fully displayed in the three-dimensional approach introduced in the section \ref{sec:3DMech}.

 All the inputs and outputs of this 4 input - 4 output port model $\mathcal{S}^{CFFF}(\mathrm{s})$ are projected in the body frame $(\mathbf x_\mathcal S,\; \mathbf y_\mathcal S,\; \mathbf z_\mathcal S)$ of the mechanism $\mathcal S$, chosen aligned with the frame of the  $\mathcal T$ mechanism (see Fig. \ref{fig:Sq_Scheme}). Using the channel inversion operation, one can also defined:
 \[
 \mathcal{S}^{CCCC}(\mathrm{s})=[\mathcal{S}^{CFFF}(\mathrm{s})]^{-1_{[7:24]}}\;.
 \]
 
\subsection{3D Mechanisms}
\label{sec:3DMech}
The two-dimensional elements introduced in Sect. \ref{sec:2DMech} and the channel inversion operation are powerful tools that can be used for the creation and assembly of complex three-dimensional structures. In the context of the multi-body approach followed in this paper, the mechanical conception focused on the definition of a unit-cube 3D module which could represent the basic building block for large space truss structures. 


\subsubsection{Cube Mechanism}
\label{sec:Cube_Mech}
\begin{figure}[htp!]
	\centering
	\begin{subfigure}{.4\textwidth}
		\includegraphics[width =1.02\textwidth]{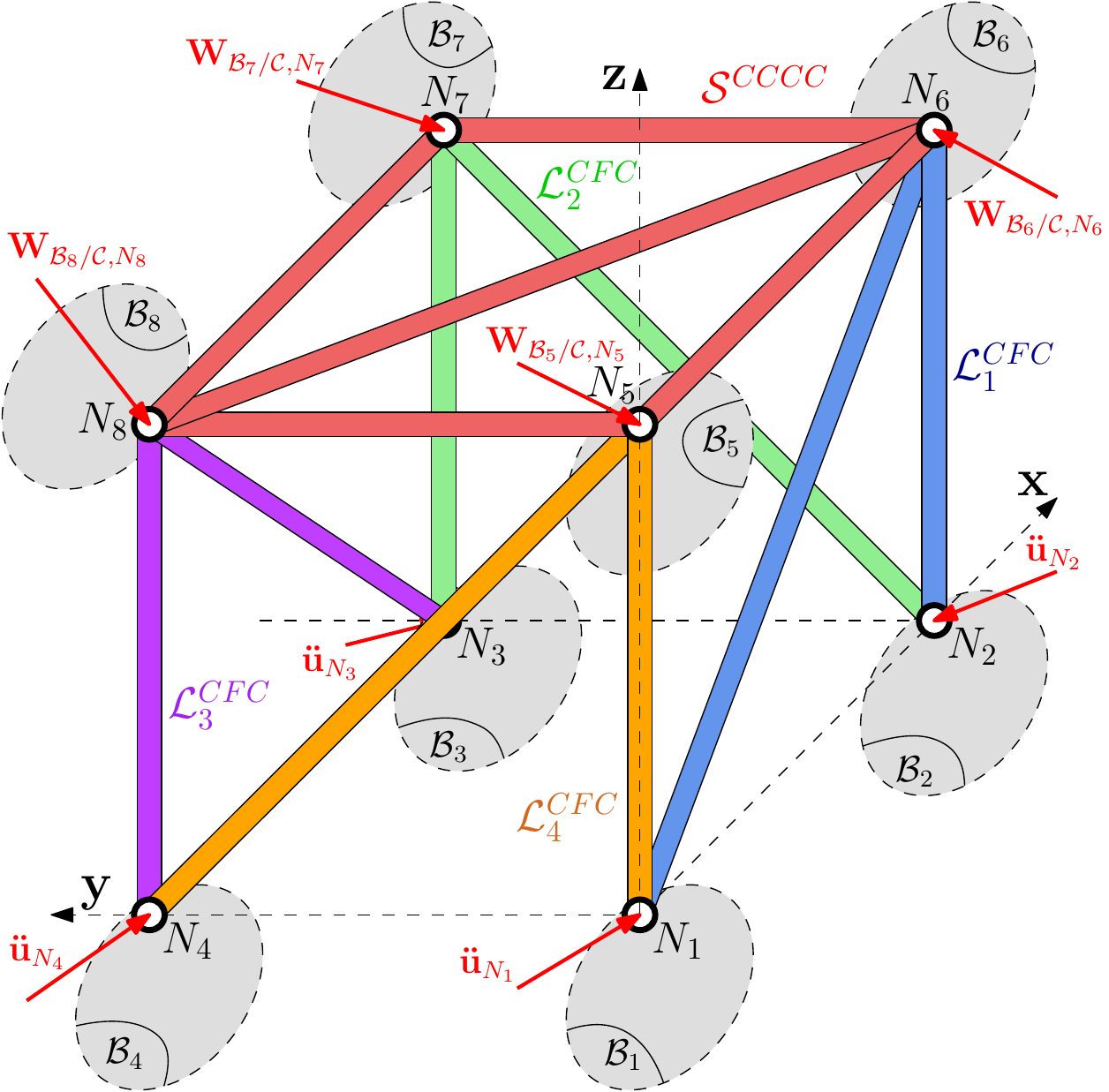}
		\caption{}
		\label{fig:Cube_Scheme}
	\end{subfigure}
	\begin{subfigure}{.59\textwidth}
		\includegraphics[width =\textwidth]{Cube_sys_block_v2}
		\caption{}
		\label{fig:Cube_Sys}
	\end{subfigure}
	\caption{Geometrical representation (a) and block diagram system (b) of the Cubic Element $\mathcal{C}(\mathrm{s})$}
	\label{fig:Cube_Notation}
\end{figure}

The Cube Mechanism $\mathcal C$ is a multi-body structure composed by 13 flexible appendages, assembled at 8 nodes to form a cubic outline with diagonal elements along the faces. A representation of its complex kinematics is given in Fig. \ref{fig:Cube_Scheme}. Each node $i$, $i=1,\cdots 8$ of the cube is connected to an external body $\mathcal B_i$ imposing an acceleration $\mathbf{\ddot u}_{N_i}$ (case of a parent body) or applying a wrench  $\mathbf W_{\mathcal B_i/\mathcal C,N_i}$ (case of child body). 

At a first glance, it can be noticed that no flexible appendix can be found to form the side of the bottom face of the Cube, along the ($x,y$) plane. This has been done to facilitate the construction of complex mechanical systems: this structure is conceived as a unit-cube-module which can be stacked on top of other elements, serially connecting them to create an elongated system. In the same fashion, multiple cubes can then be added on the sides as well. An example of this sub-structuring modeling technique will be outlined in the case study of Sect. \ref{sec:Application_Case_Study}. 

In order to facilitate this serial connection of cubes along the $z$-axis, the design of this mechanical system has been carried out to present four lower nodes (from $N_1$ to $N_4$ ) with accelerations imposed on the structure by the external parent bodies $\mathcal{B}_i  \, (i = 1,...4)$, while the upper nodes of the structure (from $N_5$ to $N_8$) are subjected to the wrenches transmitted by the external child bodies $\mathcal{B}_i \, (i = 5,...8)$. These excitation acting on the system, highlighted in red in Fig. \ref{fig:Cube_Scheme}, will represent the inputs of the 8 input - 8 output port model of the cube, denoted $\mathcal C^{CCCC-FFFF}(\mathrm s)$.  Following the general NINOP model formalism, the ouputs are the \textit{conjugate} of the inputs.

As it can be seen in Fig. \ref{fig:Cube_Sys}, the model $\mathcal C^{CCCC-FFFF}(\mathrm s)$ of this complex three-dimensional structure can be easily modeled by means of only two different 2D mechanisms: \textit{Clamped-Free-Clamped (CFC)} L-Chains $\mathcal{L}_j^{CFC}(\mathrm{s})$, repeated four times, and one single \textit{Clamped-Clamped-Clamped-Clamped (CCCC)} Square Mechanism $\mathcal{S}^{CCCC}(\mathrm{s})$. The imposed accelerations at nodes $N_1$ to $N_4$ have been inputted directly to the two \textit{clamped} ends of the L-Chains, while their \textit{free} vertex receives the combined effort of the external forces applied by $B_i$ $(i = 5,...,8)$, and $\mathcal{S}^{CCCC}(\mathrm{s})$. Exactly like in the previous cases, the feedback loops between the sub-blocks allow to satisfy the kinematics constraints and the wrench balances.

Finally, in Fig. \ref{fig:Cube_Sys} it can be noted that this model does not present DCM between the blocks. This is made possible by the fact that the changes of reference frames are handled internally within the elementary blocks, allowing the possibility to express all vectors in any generic common frame (for instance, the frame $(N_1,\mathbf x,\;\mathbf y,\;\mathbf z)$ for the cube).

\subsubsection{System Validation}

The sub-structuring technique presented in the previous sections has been implemented in \matlab/\simulink in the form of  N-Input-N-Output Port block models, which depict the dynamical behavior of each mechanism. These blocks integrate the analytical beam model of the SDT library \cite{d_alazard_f_sanfedino_satellite_2021}.

\begin{wrapfigure}[14]{r}{.4\textwidth}
	\vspace{-5pt}	
	\centering
	\includegraphics[width =\linewidth]{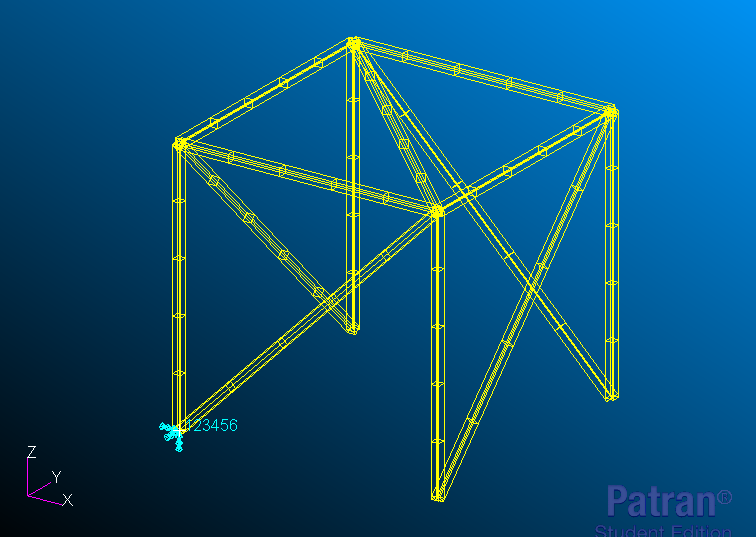}
	\caption{Validation FEM model implemented in \msc Patran}
	\label{fig:Patran_Validation}
\end{wrapfigure}
 These new models of elementary truss structures are now validated by comparison with the models obtained from a a widespread finite-element-model commercial software: \msc Patran/Nastran. This validation is performed on the Cube structure. Since it embeds the three elementary  (L-chain, Triangle and Square) sub-structures, the validation of the Cube model acts as a general validation of all its sub-structural components.

\paragraph{Geometry Definiton and Patran/Nastran Modeling}
A cube mechanism, as described in Sect. \ref{sec:Cube_Mech} is hereby considered for verification purposes. Its geometry is fixed by means of the length of its sides, $l_x$, $l_y$ and $l_z$, along the axis of the ($N_1$, $\mathbf{x}$, $\mathbf{y}$,$\mathbf{z}$) reference frame and by the mechanical characteristics of each appendage composing the kinematics. The same homogeneous beam has been repeated for each flexible body. The full mechanical characterization of the validation model is described by Table \ref{tab:verif_cube_params}, where $\nu$ is the Poisson's coefficient and $\xi$ is the damping factor.

\begin{table}[t!]
	\begin{center}
		\caption{Parameters of the cube sides and TITOP Beams used for the validation of Cube model}
		\label{tab:verif_cube_params}	
		\begin{tabular}{c c c c c c c c c c}
			\toprule
			$l_x$ [m] & $l_y$ [m] & $l_z$ [m] & $S\, [m^2]$ & $\rho \, [kg/m^2]$ & $E \, [GPa]$ & $\nu$ & $I_y \, [m^{-4}]$ & $I_z \, [m^{-4}]$ & $\xi$ \\
			\midrule
			1.0 & 1.0 & 1.0 & 9e-4 & 2700 & 70 &  0.35 & 6.75e-08 & 6.75e-08 & 0.001 	\\ \bottomrule
		\end{tabular}
		
	\end{center}
\end{table}

For the comparison with the \msc Nastran model, the cube is only clamped at node $N_1$ whose model (labeled \textit{SDT}) is obtained using the channel inversion operation:
\[
\mathcal C^{CFFF-FFFF}(\mathrm s)=[\mathcal C^{CCCC-FFFF}(\mathrm s)]^{-1_{[7:24]}}\;.
\]
This inversion allows both the comparison of the flexible mode frequencies and the modal participation factors with respect to the clamped node.

The same mechanical system has been implemented in \msc Patran, using the same mechanical characteristics of the \textit{SDT} model and modeling each beam using the \textit{CBEAM} element property, in order to take torsional behavior into account. The 3D model (labeled \textit{Patran}) created in \msc Patran is displayed in Fig. \ref{fig:Patran_Validation}, where the 5 elements used for each beam can be distinguished. The structure can be seen clamped at the origin of the ($x,y,z$) axis, which corresponds to node $N_1$.

Table \ref{tab:Verif_Cubo} describes the first ten modes of the two models (\textit{SDT} and \textit{Patran}) by means of their natural frequencies $\omega_k$ and their modal participation factors. The comparison shows a good match in both physical properties.

\begin{figure}[h!]
	\centering
	\includegraphics[width =\linewidth,trim=44 10 53 5,clip]{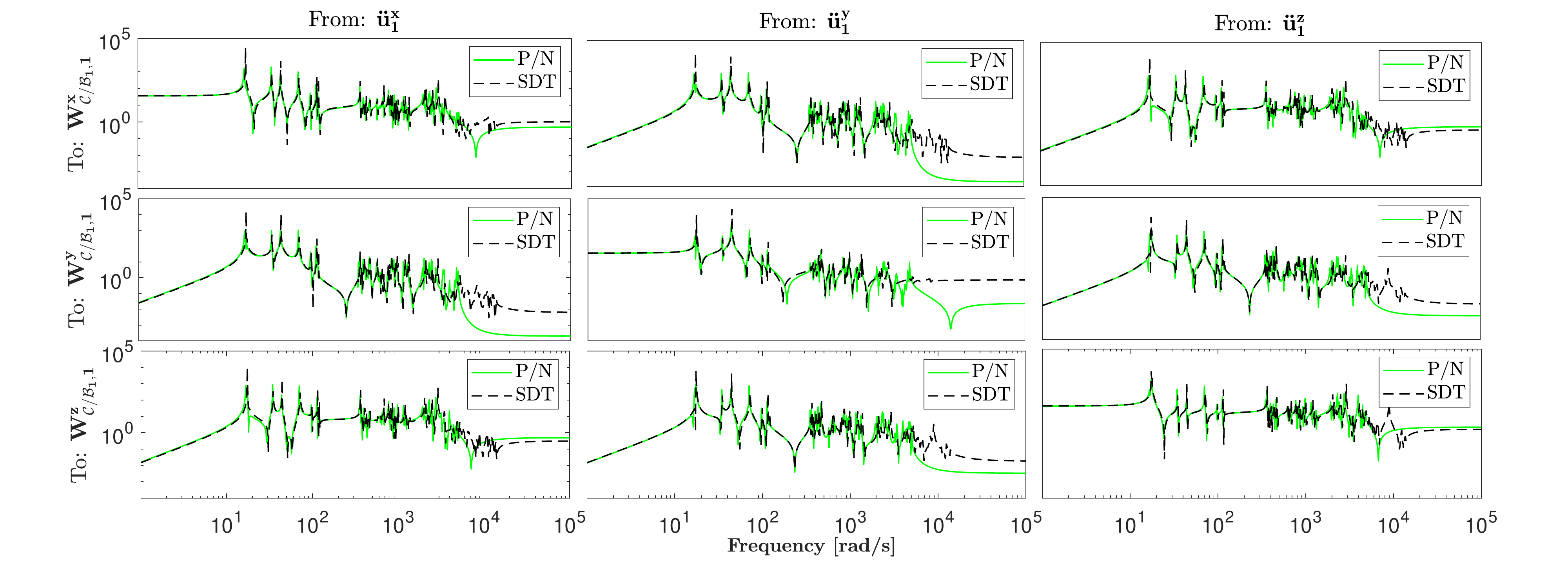}
	\caption{Verification of the Cube element's frequency response (SDT) by comparison with the \msc Patran/Nastran model (P/N)}
	\label{fig:verif_freq_resp}
\end{figure}

\begin{table}[t!]
	\caption{Comparison between the Patran/Nastran model and the SDT cube: modal frequency $\omega_k$ and participation factors ($T_1,T_2,T_3,R_1,R_2,R_3$).}
	\label{tab:Verif_Cubo}	
	\centering
	\resizebox{\textwidth}{!}{%
		\begin{tabular}{k |a  k | a  k  |a  k | a k |a k |a k |a k}
			\toprule
			\multicolumn{1}{c}{\textit{Mode}} & \multicolumn{2}{c}{$\boldsymbol{\omega_k}$ [rad/s]} & \multicolumn{2}{c}{\textit{T1}} & \multicolumn{2}{c}{\textit{T2}} & \multicolumn{2}{c}{\textit{T3}} & \multicolumn{2}{c}{R1}      & \multicolumn{2}{c}{R2}      & \multicolumn{2}{c}{R3}        \\ \midrule
			&    \textit{Patran}   & \textit{SDT}     & \textit{Patran}   & \textit{SDT}   & \textit{Patran}   & \textit{SDT}   & \textit{Patran}   & \textit{SDT}   & \textit{Patran} & \textit{SDT} & \textit{Patran} & \textit{SDT} &\textit{Patran}   & \textit{SDT} \\
			\textit{1}    & 16.77            & 17.43             & -3.830         & -3.912         & 2.040          & 2.207          & 1.714          & 1.474          & 0.113        & -0.292       & -3.746       & -3.704       & 5.127          & 5.236        \\
			\textit{2}    & 17.51            & 17.75             & -0.716         & -0.432         & 2.211          & 2.091          & -3.890         & -4.021         & -5.495       & -5.475       & 1.684        & 1.964        & 2.167          & 1.815        \\
			\textit{3}    & \textit{34.37}   & 34.90             & 2.279          & -2.271         & 1.111          & -1.256         & -0.993         & 0.921          & 0.069        & -0.018       & 4.423        & -4.315       & \textit{0.676} & -0.765       \\
			\textit{4}    & 43.66            & 44.34             & 2.066          & -2.004         & 4.546          & -4.534         & 0.749          & -0.824         & -2.424       & 2.421        & 0.031        & 0.074        & 0.503          & -0.439       \\
			\textit{5}    & 70.36            & 72.18             & 1.711          & -1.746         & 1.737          & 1.616          & 1.288          & 1.239          & -0.897       & -0.844       & -0.677       & -0.651       & -0.128         & -0.149       \\
			\textit{6}    & 97.26            & 99.54             & 0.482          & -0,428         & 0.402          & -0.389         & 0.398          & -0.392         & -0.192       & 0.187        & -0.079       & 0.086        & 0.007          & -0.002       \\
			\textit{7}    & 112.83           & 115.31            & 0.667          & -0.656         & 0.401          & -0.398         & 0.561          & -0.567         & -0.144       & 0.140        & 0.173        & -0.168       & 0.129          & -0.132       \\
			\textit{8}    & 117.91           & 120.84            & -0.554         & -0.552         & 0.067          & 0.059          & -0.074         & -0.095         & -0.026       & -0.022       & -0.005       & -0.005       & 0.043          & 0.042        \\
			\textit{9}    & 348.92           & 354.30            & -0.184         & -0.139         & 0.126          & 0.118          & 0.386          & 0.472          & -0.034       & -0.031       & -0.071       & -0.077       & 0.031          & 0.032        \\
			\textit{10}   & 361.72           & 365.24            & -0.727         & -0.761         & -0.108         & -0.123         & -1.101         & -1.078         & 0.023        & 0.027        & 0.035        & 0.026        & -0.042         & -0.047       \\ \bottomrule
		\end{tabular}%
	}
\end{table}

A more detailed validation is possible thanks to the interface between \msc Nastran  and SDT library  which allows to import the  \msc Patran model directly in \matlab/\simulink. From the Nastran \texttt{.f06} analysis file, this interface provides the $6\times 6$ transfer from the  acceleration $\mathbf{\ddot{u}}_{N_1}$ at the node $N_1$ to the reaction wrench $\mathbf{W}_{\mathcal{C}/\mathcal{B}_1,{N_1}}$ at this node. The frequency-domain response for the 3 translation degrees of freedom  of this transfer is depicted in Fig. \ref{fig:verif_freq_resp} (labeled \textit{P/N}) and compared with the ones from the proposed model  (labeled \textit{SDT}).
The analysis of these plots confirms a good match between the two models, which present the exact same steady-state response and accurate superposition of the peaks up to the high frequency bandwidth. 

\section{Integrated robust control/structure optimization of a fine pointing spacecraft}
\label{sec:Application_Case_Study}

\begin{wrapfigure}[26]{l}{.53\columnwidth}
	\vspace{-8pt}
	\centering
	\includegraphics[width =1.08\linewidth]{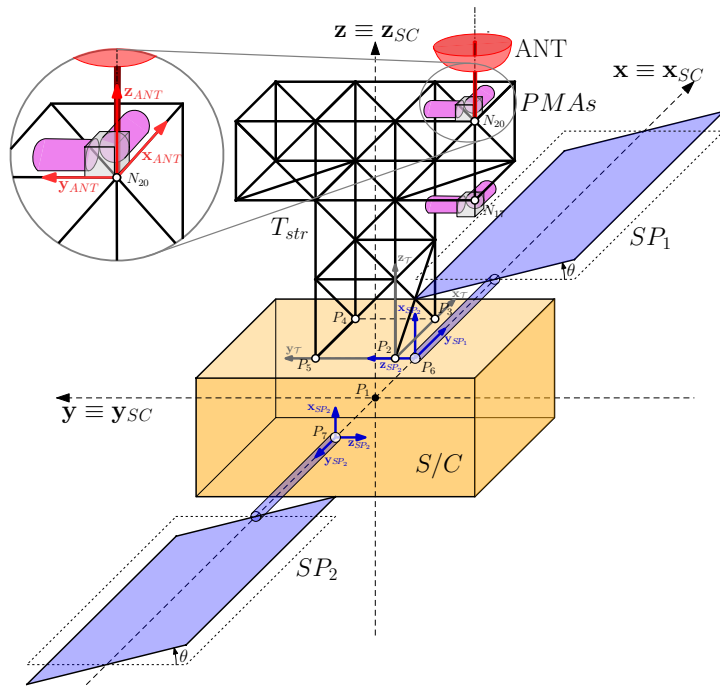}
	\caption{Overall view of satellite's case study. The spacecraft composed by a main central spacecraft body ($S/C$), two Solar Panels (${SP}_1,{SP}_2$) rotated around axis $x$ of an angle $\theta$. The HPP antenna ($ANT$) and the $4$ Proof-Mass-Actuators ($PMAs$) are both placed on top of the truss structure $T_{str}$.}
	\label{fig:total_assembly}
\end{wrapfigure}

This section introduces a study case on a space application to demonstrate the power of the proposed multi-body sub-structuring approach for modeling and control of flexible structures. 
Let us consider a Telecommunication satellite having one High-Precision-Pointing (HPP) antenna connected to the main spacecraft (S/C) body by means of a large truss structure as shown in Fig. \ref{fig:total_assembly}. Two solar arrays are connected to the sides of the satellite and are able to rotate thanks to Solar Array Driving Mechanisms (SADM), which on the other hand introduce perturbations in the attitude of the spacecraft and in the pointing of the antenna.


In this context, the blocks introduced in Sect. \ref{sec:2DMech} and \ref{sec:3DMech} have been used to construct the large T-Shaped truss structure $T_{str}$, which supports the high-precision antenna. Given that the multi-body design is based on the TITOP approach, this complex flexible system can be fully parametrized for any physical property characterizing the beams. A structural and robust control co-design has been implemented for the system, to showcase the strength of a parametric structural model by achieving the two following concurrent goals:
\begin{enumerate}
	\item Reject the perturbations introduced by the SADMs acting on the Line-of-Sight (LOS) of the Antenna by performing an active control using 4 Point-Mass-Actuators (PMAs), distributed on the structure.
	\item Minimize the mass of the T-Truss structure $T_{str}$ by reducing the section of the beams composing the system, all while complying with the pointing and vibration rejection requirements.
\end{enumerate}

The following sections will detail the design and assemblage of the T-Truss structure and the procedure chosen to implement the co-design.

\subsection{Flexible T-Truss Structure}\label{sect:T-truss}

\begin{figure}[h]
	\centering
	\includegraphics[width =\linewidth]{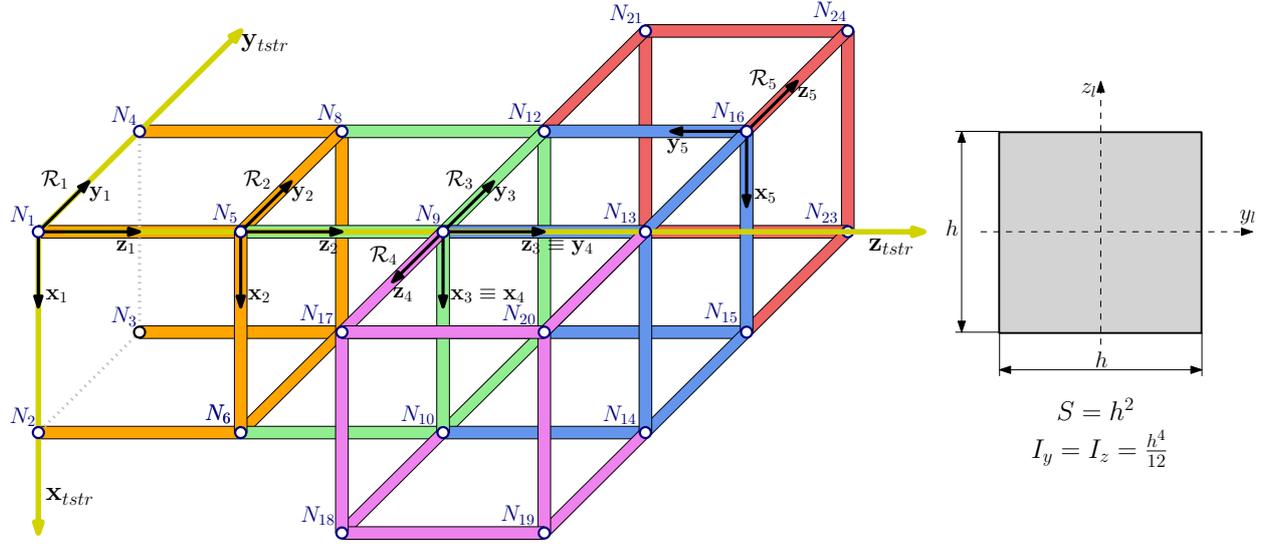}
	\caption{Simplified representation of the T-Shaped Truss Structure, without any diagonal beam in the cubes' faces, and the beams square sections. For clarity sake, the Cube elements $\mathcal{C}_i$ are represented with their local reference frames $\mathcal{R}_i$ to indicate their orientation in the truss reference frame $\mathcal{R}_{tstr}(N_1,\ \mathbf x_{tstr},\ \mathbf y_{tstr},\ \mathbf z_{tstr})$. The section area $S$ and bending inertia $I_y,I_z$ of each beam is expressed in the generic beam local reference frame $\mathcal{R}_{l}(N_l,\ x_{l},\ y_{l},\ z_{l})$ as function of the side parameter $h$.}
	\label{fig:T_struct_nom}
\end{figure}

The T-shaped Truss antenna support has been modeled by means of five Cube mechanism elements, connected to each other to form the structure seen in Fig. \ref{fig:T_struct_nom}, where the diagonal beams on the faces have been hidden to facilitate reader's understanding of the architecture. Overall, this complex mechanical system is composed by 65 flexible beams. Among them, it is possible to identify a core assembly of three cubes ($\mathcal{C}_1$, $\mathcal{C}_2$ and $\mathcal{C}_3$) which are simply serially stacked on top of each other, along the $\mathbf z_{tstr}$ axis of the spacecraft reference frame $\mathcal{R}_{tstr}(N_1$, $\mathbf x_{tstr},\mathbf y_{tstr},\mathbf z_{tstr})$. The remaining two cubes are then placed on the sides of the previous assembly, developing along the $\mathbf y_{tstr}$ axis direction. 
This is achieved by the multiple use of two direction cosine matrices, which perform a rotation of the cubes around the $\mathbf x_{tstr}$ axis. This results in cube $\mathcal{C}_4$ facing towards the negative direction of $\mathbf y_{tstr}$ while $\mathcal{C}_5$ towards the positive one.

\begin{table}[t!]
	\begin{center}
		\caption{Parameters of TITOP Beams used for the T-Truss Structure}
		\label{tab:GeomParam}	
		\begin{tabular}{c c c c c c c}
			\toprule
			$S\, [m^2]$ & $\rho \, [kg/m^2]$ & $E \, [GPa]$ & $\nu$ & $I_y \, [m^{-4}]$ & $I_z \, [m^{-4}]$ & $\xi$ \\
			\midrule
			${h}^2$ & 2700 & 70 &  0.35 & $h^4 /12 $ & $h^4 /12 $ & 0.001 	\\ \bottomrule
		\end{tabular}
		
	\end{center}
\end{table}

The overall structure is composed by 24 nodes in total. Four of them - nodes from $N_1$ to $N_4$ - are connected to the S/C main body and therefore are designed to have an acceleration imposed onto them. The rest of the points on the other hand can receive an external input in the form of an external wrench. 

This is the case for nodes $N_{17}$ and $N_{20}$, where the PMAs and the antenna is connected, while no external forcing term are applied to other points of the structure.
In terms of physical and geometrical characterization of the system, it has a total envelope of $1m \times 3m \times 3 m $, with each single cube having a volume of $1m \times 1m \times 1m$. The same beam properties have been repeated for all the 65 beams composing the structure. Given an aluminum beam with a square section, a parametric model has been implemented by means of the section length $h$. This variable drives both the section area and the second moments of area of the beam, as displayed in Fig. \ref{fig:T_struct_nom}. The parameter will then be at the center of the structural-control co-design of section \ref{sec:Application_Case_Study}, as it is directly related to both mass and stiffness properties of the system. Finally, the mechanical characterization of all the beams is given in Table \ref{tab:GeomParam}.

%
%

\subsection{Proof Mass Actuators (PMAs) Model}\label{sec:PMA}
\begin{wrapfigure}[9]{r}{.45\textwidth}
	\centering
	\vspace{-20pt}
	\includegraphics[width=\linewidth]{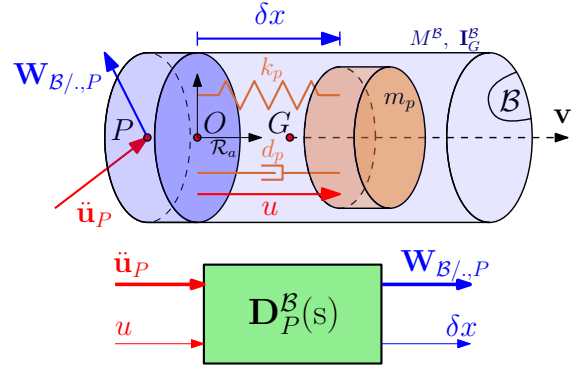}
	\caption{Proof Mass Actuator (PMA) mechanical system and its associated block diagram system $\mathbf{D}_P^{\mathcal{B}}(\mathrm{s})$ in the TITOP approach}
	\label{fig:PMA}
\end{wrapfigure}

The Proof Mass Actuator mechanical system has been modeled in Fig. \ref{fig:PMA} as the body $\mathcal{B}$, composed by a rigid casing and a one-dimensional spring-mass-damper system. Under these assumptions, the mechanical actuator is fully defined by the following set of parameters:


\begin{itemize}
	\item $\mathcal{R}_a$: local reference frame attached to the PMA at the reference point $O$;
	\item $\mathbf v$: the unit vector along the PMA axis, expressed in $\mathcal{R}_a$;
	\item $G$ and $P$: the center of mass of the PMA (at rest) and the connection point to the parent body, respectively;
	\item $M^\mathcal{B}$ and $\mathbf{I}_G^\mathcal{B}$: respectively, the mass and the inertia matrix at $G$ of the PMA casing;
	\item $m_p$, $k_p$, $d_p$.: the mass, stiffness and damper of the spring-mass-damper system describing the dynamics of the PMA along the axis $\mathbf{v}$;
	\item $u$: control effort applied on the proof mass along the axis $\mathbf{v}$;
	\item $\delta_x$. Relative displacement of the proof mass with respect to the casing;
	\item $\mathbf{\ddot{u}}_P$:  $\{6 \times 1 \}$ acceleration twist  of the PMA at the connection point expressed in the $\mathcal{R}_a$ frame;
	\item $\mathbf{W}_{\mathcal{B}/.,P}$:  $\{6 \times 1 \}$ wrench applied by the PMA at point $P$, expressed in the  $\mathcal{R}_a$ frame;
\end{itemize}

The Proof Mass Actuator mechanical system $\mathcal{B}$ has then been modeled as $\mathbf{D}^{\mathcal{B}}_P(\mathrm{s})$, a $\{7 \times 7 \}$ linear dynamic model in the TITOP approach. The system is obtained by means of the governing equations of the mechanical dynamics, defined as follows:
\begin{align}
&m([\mathbf{v}^T \quad \mathbf{0}_{1\times 3}]\mathbf{\ddot{u}}_G + \ddot{\delta}_x ) = -k \delta_x -d \dot{\delta}_x+ u \\
&\mathbf{\ddot{u}}_G = \boldsymbol{\tau_{GP}} \mathbf{\ddot{u}}_P \\
&\mathbf{W}_{\mathcal{B}/.,P} = -\boldsymbol{\tau_{GP}}^T \left( \begin{bmatrix}
M \mathbf{I}_{3} & \mathbf{0}_3 \\
\mathbf{0}_3 & \mathbf{I}^{\mathcal{B}}_G \\
\end{bmatrix}\boldsymbol{\tau_{GP}}\mathbf{\ddot{u}}_P + m \begin{bmatrix}
\mathbf{v} \\ \mathbf{0} \\
\end{bmatrix} \ddot{\delta_x} \right) \\
&\boldsymbol{\tau_{GP}} = \begin{bmatrix}
\mathbf{I}_3 & [\mathbf{r_{GP}}]_\times \\
\mathbf{0}_{3\times 3} & \mathbf{I}_3 \\
\end{bmatrix}
\end{align}
where $\boldsymbol{{\tau}}_{GP}$ is the \textit{Kinematic Model} of the rigid link between point $G$ and point $P$, defined thanks to the skew matrix $[\mathbf{r}_{GP}]_\times$ associated to the vector $\mathbf{r}_{GP}$ from node $G$ to node $P$.

The PMA $7\times 7$  block-diagram model $\mathbf{D}^{\mathcal{B}}_P(\mathrm{s})$ is showcased in Fig. \ref{fig:PMA}. The first $6\times 6$  port of the model defines the transfer between the acceleration $\mathbf{\ddot{u}}_P$ and the wrench $\mathbf{W}_{\mathcal{B}/.,P}$, while the last describes the one between the control effort $u$ and the proof mass relative displacement $\delta_x$. 
The mechanical parameters which describe the PMAs used in the current case study are given in appendix, Table \ref{tab:Sat_prop}.
%


\subsection{System Modeling}
\subsubsection{Full order Linear Parameter Varying (LPV) model}
\label{sec:OL_sys}
\begin{figure}[h]
	\centering
	\includegraphics[width =\linewidth]{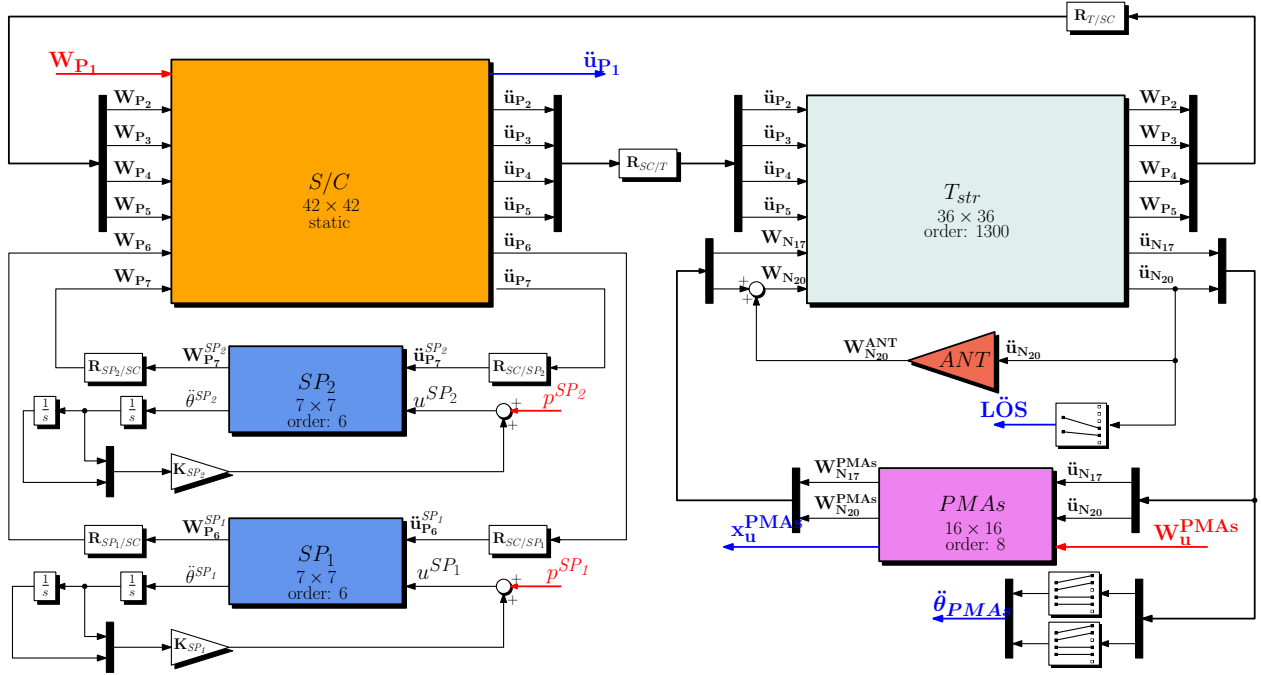}
	\caption{Open Loop System of the spacecraft in Fig. \ref{fig:total_assembly}.}
	\label{fig:Linear_OL_system}
\end{figure}

The complete system in Fig. \ref{fig:total_assembly} is modeled by means of a block-diagram approach using elements derived from the SDT library introduced in \cite{d_alazard_f_sanfedino_satellite_2021}.  The block-diagram representation of the whole spacecraft is then depicted in Fig. \ref{fig:Linear_OL_system} and is detailed by:

%

\begin{itemize}
	\item \textit{Central body ($S/C$)}. This block is a static $42 \times 42$ multi-port rigid body model.  It models the dynamics of a rigid body  subjected to multiple wrenches applied at connection points $P_1$, the center of mass where is located the Attitude Control System (ACS),  $P_2$, $P_3$, $P_4$, $P_5$ the $4$ connection points with the T-truss structure, and $P_6$, $P_7$ the connection points with the $2$ solar panels. The geometry of these points is detailed in Appendix, Table \ref{tab:Connect_points}.
	
	\item \textit{Solar Panels ($S\!P_i,\;i=1,2$)}. These $2$  blocks are $6$-th order  single port flexible body models. Each solar panel model takes into account $3$ flexible modes and  is connected to the  central body by a revolute joint driven along the $x$-axis by the SADM. The two solar panels are identical. The joint angular configuration $\theta$ is taken into account in the DCM $R_{SC/SP_i}$. The SADM is the main source of disturbance and is modeled by a local stiffness $k_{S\!P_i} $, a viscous friction $f_{SP_i}$ and an internal disturbing torque $p^{S\!P_i}$. The overall torque acting on the mechanisms is:	
	\begin{equation}
	u^{SP_i} = k_{S\!P_i} \delta\theta_i + f_{S\!P_i}\delta\dot{\theta}_i + p^{S\!P_i}
	\end{equation}
	where $\delta\theta_i$ and $\delta\dot{\theta}_i$ are obtained integrating the relative acceleration $\delta\ddot{\theta}_i$ of $S\!P_i$ with respect to $S/C$. The two coefficients $k_{S\!P_i}$, $f_{S\!P_i}$ are contained in the matrices: $\mathbf{K}_{S\!P_i} = [f_{S\!P_i}, k_{S\!P_i}]$.
	
	\item \textit{Antenna ($ANT$)}. The antenna is modeled as a rigid body described by its mass $M_{ANT}$ and its inertia matrix $I_{ANT}$  at the node $N_{20}$ in its local frame aligned with the S/C reference frame.  The LOS  direction of the antenna therefore coincides with the $z$-axis of the spacecraft reference frame $\mathcal{R}_{SC} (P_1,\ \mathbf x_{SC},\ \mathbf y_{SC},\ \mathbf z_{SC} )$.
	
	\item \textit{T-truss structure ($T_{str}$)}. This $36\times 36$ block is the $6$ input - $6$ output port model described in section \ref{sect:T-truss}. The $6$ connection points are $N_1$, $N_2$, $N_3$, $N_4$  (with the parent central body), $N_{17}$ and $N_{20}$ (with the child antenna and PMAs). Since it is composed of $65$ flexible beams and since the SDT analytical model of a beam is a $20$-th order model, this block is a $1300$-th order model.
	
	\item \textit{Proof Mass Actuators ($PMAs$)}. This $16\times 16$, $8$-th order block embeds the  $4$ PMAs as described in \ref{sec:PMA}.  The positioning of these PMAs at node $N_{20}$ and $N_{17}$ has been selected to maximize the control effect over the orientation of the LOS. Thanks to the use of two couples of PMAs acting in the same direction, the control system is able to control the elevation and the azimuth of the LOS and the $2$ translations in the plane orthogonal to the LOS. 
	
\end{itemize}



The data of all spacecraft sub-components can be found in Appendix, Table \ref{tab:Sat_prop}.

The full order model of the whole system is thus a $1324$-th order model. Its $20$ outputs $\mathbf y$, seen in blue in Fig. \ref{fig:Linear_OL_system} are:
\begin{itemize}
	\item $\mathbf{\ddot{LOS}}$:  $2\times 1$ vector with LOS elevation and azimuth angular accelerations. This is one of the parameters taken into account in the performance index of the co-design,
	\item $\mathbf{x_u^{PMAs}}$:  $4\times 1$ vector of the linear internal displacements of the 4 PMAs.
	\item  $\mathbf{\ddot{u}_{P_1}}$: $6\times 1$ acceleration vector of the center of mass of the central body, 
	\item $\boldsymbol{\ddot{\theta}_{PMA_s}}$: $8\times 1$ vector of  the linear and angular accelerations at nodes $N_{17}$ and $N_{20}$ along and around the $\mathbf x$ and $\mathbf y$ axes.
\end{itemize}
The  $12$ inputs $\mathbf u$ of the model, highlighted in red in Fig. \ref{fig:Linear_OL_system}, are:
\begin{itemize}
	\item $p^{SP_1},\, p^{S\!P_2}$: the two internal perturbation torques acting at the revolute joints of the each SADM,
	\item $\mathbf{W_{P_1}}$: $6\times 1$ wrench vector at the center of mass of the spacecraft $P_1$,
	\item $\mathbf{W_u^{PMAs}}$:  $4 \times 1$ vector of linear forces applied to the system at nodes $N_{17}$ and $N_{20}$ by the four PMAs.
\end{itemize}
One of the main interest of the SDT lies in the possibility of declaring the various mechanical and geometrical parameters as varying parameters or uncertain parameters. Thus, the LPV model directly computed from the block-diagram description is fully compatible with the \matlab robust control toolbox to perform robust performance analyzes and robust control designs. That is particularly relevant during preliminary design  of Space systems to face unpredictable manufacturing inaccuracies and also uncertainties or misknowledge on some sub-systems whose the design is not finalized. It is all the more relevant for preliminary co-design where some of these parameters are tunable parameters. In this case study, two kinds of parameters are considered:
\begin{itemize}
	\item the \textbf{uncertain parameters}:  $\boldsymbol \delta=[\delta_{M_{S/C}},\ \delta_{I_{yy,\,SC}}, \ \delta_{\omega_{1,_{S\!P}}},\ \tau]^T$ composed of normalized ($\in[-1,\;1]$) uncertainties on the mass of the main spacecraft hub $M_{SC}$, the inertia along the $y$-axis of the same body $I_{yy_{SC}}$, the first natural frequency of both solar panels $\omega_{1_{{S\!P}}}$ and the angular configuration $\theta\in[-\pi,\;±\pi]$ of the solar panels parametrized by $\tau = \tan(\theta/4) \in [-1,1]$ as introduced in \cite{Guy2014, dubanchet2016modeling}.  The uncertainty values are defined in Appendix, Table  \ref{tab:Sat_prop},
	\item the \textbf{tunable parameter}: $\boldsymbol\Theta=h$, the variation on the side of the T-truss beam section such that:
	\begin{equation}
		h \in \mathcal{H} = [1.5\, cm , 3 \, cm]\;.
	\end{equation}
\end{itemize}
From the block-diagram model depicted in Fig. \ref{fig:total_assembly}, the LPV full order model $\mathcal{G}(\boldsymbol{\Theta,\delta})$ is directly computed thanks to the \matlab function \texttt{ulinearize} and fully characterized by the arguments of the following Linear  Fractional Transformation (LFT):
\begin{equation}
	\mathcal{G}(\boldsymbol{\Theta,\Delta})=\mathcal F_u(\mathcal{G}(\mathbf{\Theta}),\boldsymbol\Delta)\quad\mbox{and}\quad \mathcal{G}(\mathbf{\Theta})=\mathcal F_u(\mathbf G(\mathrm s),\boldsymbol{\Delta_\Theta})
\end{equation}
with $\boldsymbol \Delta$ the uncertainty parameter block and $\boldsymbol{\Delta_\Theta}$ the tunable parameter block defined by:
%
%
\begin{align}
	\boldsymbol{\Delta} = \text{diag} \big( \boldsymbol{\Delta}_\tau, \mathbf{\Delta}_p \big) ,
	\quad \boldsymbol{\Delta}_\tau = {\tau}\ \mathbf{I_{16}} ,
	\quad \mathbf{\Delta}_p =\text{diag} \big( \delta_{M_{S/C}} \mathbf{I_3},  \delta_{I_{yy}} ,\delta_{\omega_{1,S\!P}}\ \mathbf{I_4}\big),
\quad \boldsymbol{\Delta_\Theta}=h\,  \mathbf{I_{905}}
\end{align}
where $\mathbf I_n$ is the identity matrix of size $n$. This size indicates the number of occurrences of each parameters. Thus the system presents an extremely high number of occurrences for the tunable parameter $h$, amounting to 905 occurrences.

The upper LFT $\mathcal F_u(.,.)$ can be interpreted by a block-diagram operation as depicted in Fig. \ref{fig:closed_loop_sys} for the reduced order model $\mathcal{G}_r(\mathbf{\Theta})$ presented in the next section.

\subsubsection{System Reduction}
\begin{figure}[h!]
	\centering
	\includegraphics[width =\columnwidth]{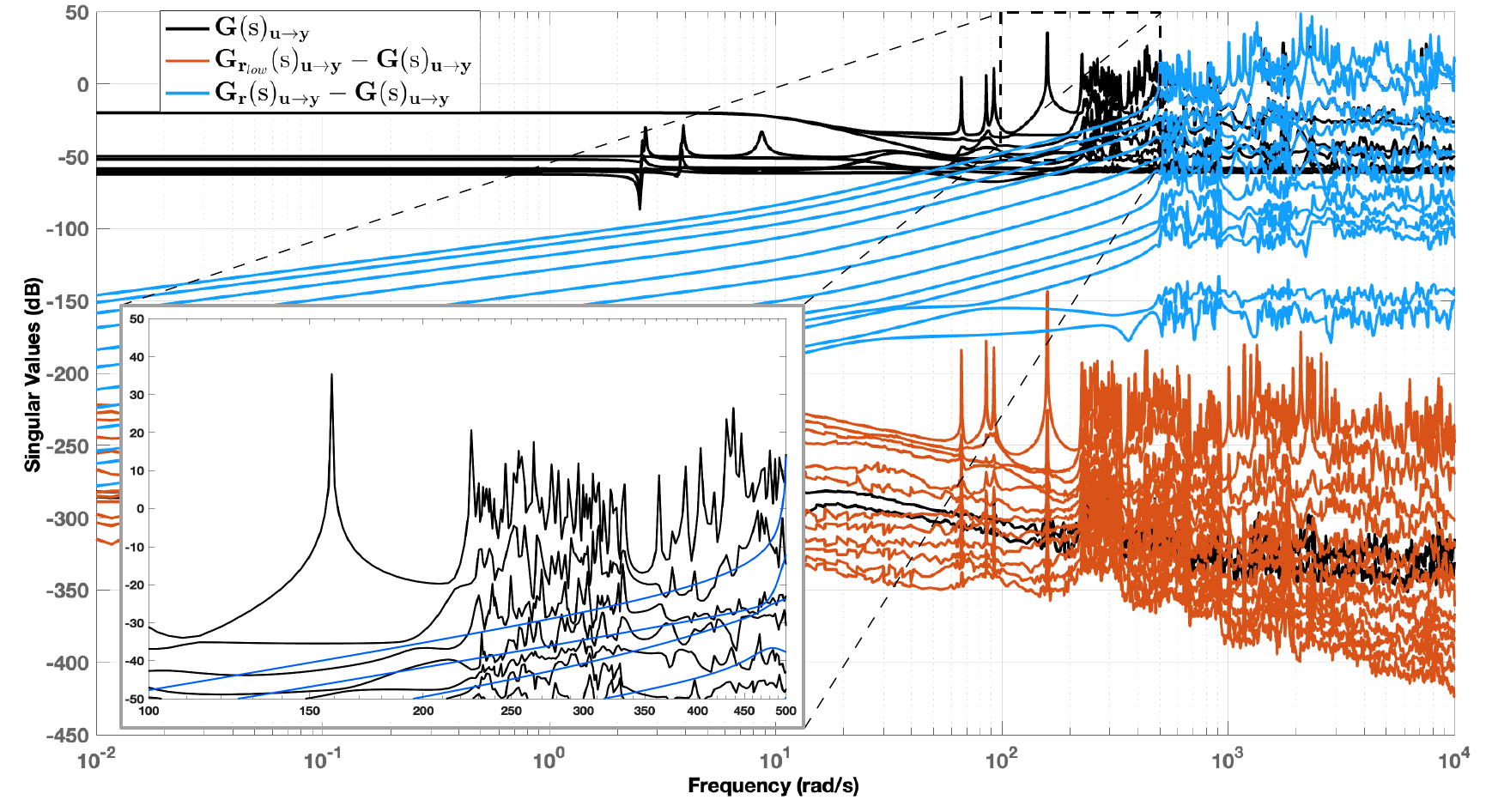}
	\caption{Comparison on the input-output transfer singular value responses between the full order model $\mathbf{G}_{r}(s)$, and the reduced order models $\mathbf{G_r}_{low}(s)$, truncated at low frequencies, and $\mathbf{G}_{r}(s)$, truncated both at low and high frequencies.}
	\label{fig:sys_comp_reduct}
\end{figure}

As mentioned in the previous section, the order of  $\mathbf G(\mathrm s)$ is very high ($1324)$. A reduction in the modal state-space representation of $\mathbf G(\mathrm s)$ is thus performed to remove the very slow poles and the high frequency flexible modes based on the following considerations:
\begin{itemize}
	\item  in the modeling of the T-truss structure, the numerous loop closure constraint are solved thanks to the channel inversion operation. Each closed-loop kinematic chain concerns the $6$ d.o.f.s and leads to $12$ poles at $0$ frequency. But numerical errors during the inversion make that these poles are not exactly at null frequency. These very slow poles do not contribute in the input-output transfer of $\mathbf G(\mathrm s)$ and they can be removed by a reduction in the modal realization of $\mathbf G(\mathrm s)$ of all the stable poles with a magnitude lower than $0.01\,rad/s$. This reduction removes $540$ ($=45\times 12$ considering the $45$ closed-loop kinematic chains inside the T-truss structure)  poles. This reduced order model is denoted $\mathbf{G_r}_{{low}}(\mathrm{s})$,
	\item very high frequency flexible mode, with a frequency greater than $500\,rad/s$, are out of the frequency range considered for the active control with the PMAs. They are also reduced in the modal realization of $\mathbf G(\mathrm s)$. This final reduced order model is denoted $\mathbf{G_r}(\mathrm{s})$ and its order is $178$. 
\end{itemize}
The goodness of the reduction can be appreciated by comparing the frequency-domain responses of the input-output transfer ($\mathbf u \to \mathbf y$) between the full order model  $\mathbf{G}(\mathrm{s})$ and the 2 reduced order models $\mathbf{G_r}_{low}(s)$ and $\mathbf{G_r}(\mathrm{s})$, as done in Fig. \ref{fig:sys_comp_reduct}. The magnitude of $\mathbf G(\mathrm s)-\mathbf{G_r}_{{low}}(\mathrm{s})$ is totally neglectable, showing the irrelevance of the loop-closure poles, while the magnitude $\mathbf G(\mathrm s)-\mathbf{G_r}(\mathrm{s})$ shows an  acceptable error. The final model $\mathbf{G}_r(s)$ correctly represents the dynamics of the open-loop system in the frequency domain of interest and will be used to perform all further analysis and controller synthesis using the upper linear fractional transformations $\mathcal{G}_r(\mathbf{\Theta}) = \mathcal{F}_u (\mathbf{G_r}(\mathrm{s}), \mathbf{\Theta})$ for the LPV system and $\mathcal{F}_u (\mathcal{G}_r(\mathbf{\Theta}), \boldsymbol{\Delta} )$ for the overall uncertain open-loop system. 

Nevertheless, the performance robustness analysis proposed in section \ref{sec:mu} will consider the model $\mathbf{G_r}_{{low}}(\mathrm{s})$ with all the flexible modes.

\subsection{Closed-Loop System}
\label{sec:CL_sys}
\begin{figure}[h]
 	\centering
 	\includegraphics[width =.8\linewidth]{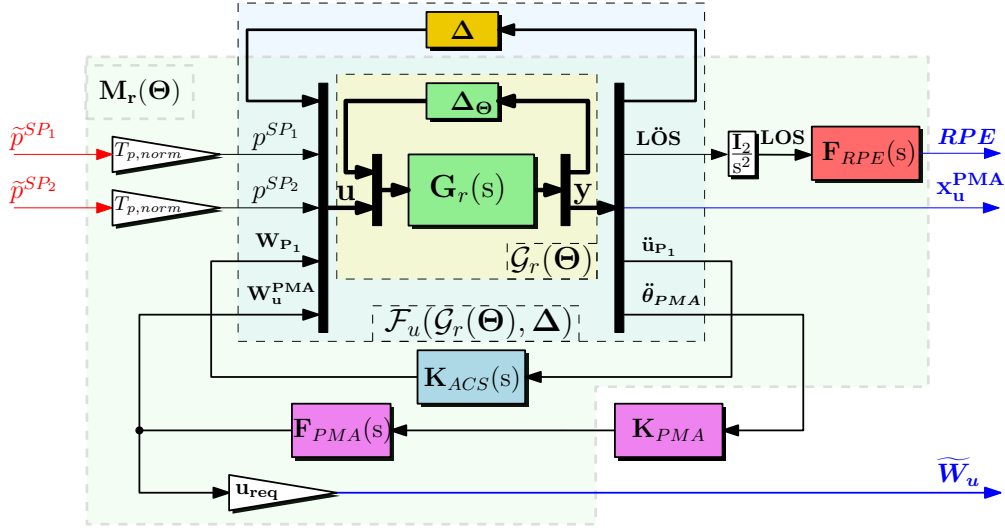}
 	\caption{Closed Loop System for Synthesis}
 	\label{fig:closed_loop_sys}
 \end{figure}

As previously stated, the main objective of this case study is to demonstrate the capability of performing structural and control co-design using parametric models for the flexible appendages, suited for robust analysis. In this context, the reduced open-loop linear model described in the previous section can augmented with the controllers acting on the system and the pointing performance indexes as shown in Fig. \ref{fig:closed_loop_sys}.

\subsubsection{Control system architecture}
The closed loop system is composed by two controllers:
\begin{itemize}
	\item  \textit{The Attitude Control System $ACS$}. The ACS implemented for this mission is a basic proportional-derivative (PD) controller aimed at fixing the attitude of the satellite in space. 
		\begin{equation}
	\mathbf{K}_{ACS}(\mathrm s) = -[\mathbf 0_{3\times 3}\quad \mathbf I_3]^T \frac{\mathbf{k}_p+\mathbf{k}_v\,\mathrm s}{\mathrm s^2} [\mathbf 0_{3\times 3}\quad \mathbf I_3]
	\end{equation}	
	$\mathbf{K}_{ACS}(\mathrm s)$ includes the rigid motion of the whole spacecraft. The decentralized (diagonal) controller gains have been chosen in order to fix the poles of the ACS at a frequency of $\omega_{ACS} = 0.1$ rad/s with a damping of $\xi_{ACS} = 0.7$ assuming the whole spacecraft is rigid:
	\begin{align}
		\mathbf{k}_p = \omega_{ACS}^2\mbox{diag} (\mathbf{J}_{tot}), & \qquad \mathbf{k}_v = 2 \xi_{ACS} \omega_{ACS}\mbox{diag}( \mathbf{J}_{tot})
	\end{align}
	where $\mathbf{J}_{tot}$ is the total inertia matrix of the satellite computed at point $P_1$. 

	
	\item \textit{PMAs Controller}. The controller for the actuator was implemented as by means of the filter $\mathbf{F_{PMAs}}(\mathrm s)$ and a static gain given by matrix $\mathbf{K_{PMA}}$. $\mathbf{F_{PMAs}}(\mathrm s)$  is a diagonal $4\times 4$ filter composed of an integrator, required to estimate the velocity from the measured acceleration, and a wash-out filter with a cut-off frequency  $\omega_{PMA} = 5\omega_{ACS} = 0.5$ rad/s:
	\begin{equation}
		\mathbf{F}_{PMA}(\mathrm{s}) = \frac{\mathrm{s}}{\mathrm{s}^2 + 2 \omega_{PMA} \xi_{PMA} \mathrm{s} + \omega_{PMA}^2} \mathbf{I_{4}}\;.
	\end{equation}
\end{itemize}

\subsubsection{Fine pointing performance definition}

The fine pointing performance is expressed in form of Relative Pointing Error (RPE) as defined in \cite{ott2011esa}. The RPE index represents the statistics of the instantaneous angular difference between the pointing vector $\mathbf{LOS}$ and its short-time average in a time interval $t_\Delta$. In the frequency domain this filter corresponds to an high-pass weight \cite{ott2013precision} applied to the system:
\begin{equation}
\mathbf{F}_{RPE}(\mathrm{s})=\epsilon_{max}^{-1} \frac{ t_\Delta \mathrm{s} \, (t_\Delta \mathrm{s} +\sqrt{12})}{( t_\Delta \mathrm{s})^{2}+6(t_\Delta \mathrm{s})+12}\mathbf{I_2}
\end{equation}

where $\epsilon_{max}$ is the maximum RPE deviation allowed, which bounds the transfer of the filter. For the present study case the following values were selected: $\epsilon_{max} = 50 \, \mu rad $, $t_\Delta = 3\, ms$.

Moreover, the SADM perturbation inputs on the system have been normalized by an estimated maximum torque acting on the system, indicated by the upper bound $p_{max}$: $T_{p,norm} = p_{max} = 0.3820 \ Nm$.
Finally, a bound for the effort generated by the PMAs on the structure has been introduced. This represents a physical constraint in the actuators' capabilities of controlling the system and it has been set at $\mathbf{u}_{req} = u_{max}^{-1} \mathbf{I_4}$, with $u_{max} = 15 \, N$.

By assembling all these blocks, the closed-loop LPV system $\mathbf{M_r}(\mathbf{\Theta})$ displayed in Fig. \ref{fig:closed_loop_sys} is obtained. The uncertainties are added to the model by means of the LFT form $\mathcal{F}_u (\mathbf{M_r}(\mathbf{\Theta}), \boldsymbol{\Delta})$. The model can then be augmented with the PMAs controller gains contained in the  $\{4\times 8\}$ scalar matrix $\mathbf{K_{PMA}}$ by means of a lower Linear Fractional Transformation $\mathcal{F}_l\Big(\,\mathcal{F}_u \big(\mathbf{M_r}(\mathbf{\Theta}), \boldsymbol{\Delta}\big), \, \mathbf{K_{PMA}}\Big)$.  The selection of the proper gains to drive the actuation of the PMAs will be the main focus of the structural and control co-design of the section \ref{sec:optimization}.




	



\subsection{Co-design optimization architecture}
\label{sec:optimization}
The objective of this study case is to implement a co-design of both the $T_{str}$ structure and control laws of the PMAs to maximize pointing performances and minimize the mass of the system at the same time. For our study, this means reducing the transfer between the normalized SADM perturbation torques $\mathbf{\widetilde{p}} = [\widetilde{p}^{SP_1}, \widetilde{p}^{SP_1}]$ and the $RPE$ index below a certain requirement. In addition, the system has limited actuation and a fixed order controller, which act as design constraints.  The co-design can therefore be classified by a non-convex optimization problem with multiple constraints, formally described as follows.

Let us consider the LFT closed-loop system $\mathcal{F}_l\Big(\,\mathcal{F}_u \big(\mathbf{M_r}(\mathbf{\Theta}), \boldsymbol{\Delta}\big), \, \mathbf{K_{PMA}}\Big)$. Thus the robust performance index to be minimized is relative to the worst-case transfer between the normalized perturbations $\mathbf{\widetilde{p}}$ and the pointing index $\mathbf{RPE}$ is defined as:
\begin{equation}
	J_c \,(\boldsymbol{\Theta}, \mathbf{K_{PMA}}) = \underset{\boldsymbol{\Delta}}{\mathrm{max}}\Big\lVert \mathcal{F}_l\Big(\,\mathcal{F}_u \big(\mathbf{M_r}(\mathbf{\Theta}), \boldsymbol{\Delta}\big), \, \mathbf{K_{PMA}}\Big)_{\mathbf{\widetilde{p}} \rightarrow \mathbf{RPE}}  \Big\rVert_\infty
\end{equation}

As previously stated, the design parameter $\boldsymbol{\Theta}$ chosen to drive the mechanical design is the side $h$ of the cross-section of each beam of the $T_{str}$ structure. Reducing this variable means lowering both mass and stiffness of the structure.
Multiple constraints are applied to this system. The worst-case maximum effort generated by the PMAs is the following:
\begin{equation}
	\mathcal{Z}_c \,(\boldsymbol{\Theta}, \mathbf{K_{PMA}}) = \underset{\boldsymbol{\Delta}}{\mathrm{max}}	\big\lVert \mathcal{F}_l\Big(\,\mathcal{F}_u \big(\mathbf{M_r}(\mathbf{\Theta}), \boldsymbol{\Delta}\big), \, \mathbf{K_{PMA}}\Big)_{\mathbf{\widetilde{p}} \rightarrow \mathbf{\widetilde{W}}_u } \big\rVert_\infty
\end{equation}  
which is bounded by the scalar constraint $\gamma_{u}^{PMA}$. Moreover, the mechanical design parameter $\mathbf{\Theta}$ belongs to a finite interval, $\mathcal{H} = [0.015, 0.03] \,m$, and the controller has a fixed static gain size of $\{4\times8\}$ defined by the family of matrices $\mathcal{K}$.

The non-smooth robust optimization problem can therefore be formulated as:

\begin{equation}
	\underset{\mathbf{K_{PMA}}, \mathbf{\Theta}}{\text{min}} \,\,\underset{\boldsymbol{\Delta}}{\mathrm{max}} \,\, f ( \mathbf{\Theta},\ \mathbf{{K}_{PMA}},\ \boldsymbol{\Delta}) \quad s.t. \;
	\begin{cases}
	\mathcal{Z}_c < \gamma_{u}^{PMA} \\
	\mathbf{K_{PMAs}} \in \mathcal{K} \\
	\mathbf{\Theta} \in \mathcal{H} \\
	\end{cases}
\end{equation}

where the minimization is performed on the objective function $f$, which depends on both $J_c$ and $\boldsymbol{\Theta}$ by means of the optimization variables ($\mathbf{K_{PMA}}$, $\boldsymbol{\Theta}$) and it is subjected to the parametric uncertainties $\boldsymbol{\Delta}$. 
This objective function $f$ can be seen as the combination of a structural $f_S$ and a control $f_C$ sub-optimization problem. The first depends only on the mass of the mechanical system, while the latter is defined as a function of $J_c$.

Since optimizing the structural sub-problem corresponds to minimize the mass of the system, it can be remarked that $f_S$ is only function of $\boldsymbol{\Theta}$. On the other hand, $f_C$ and therefore $J_c$ depend on both structural and control parameters. This implies that the coupling between the optimization problem and the control one is unidirectional \cite{frischknecht2011pareto}. 

The overall objective function for the co-design $f$ can therefore be seen as the linear combination of the two following sub-problems:
	
	\begin{equation}
	\label{eq:generic_optim_funct}
		f(\boldsymbol{\Theta},\mathbf{K_{PMA}}) = w_S f_S(\boldsymbol{\Theta}) + w_C f_C(\boldsymbol{\Theta}, \mathbf{K_{PMA}})
	\end{equation}

where $w_S$ and $w_C$ are weighting coefficients. The resolution of this non-trivial optimization problem has been performed numerically by means of a nested optimization routine, as proposed by \cite{fathy2001coupling,reyer2001comparison}. For this optimization method, the solution for the overall optimization problem is found with respect to $\boldsymbol{\Theta}$, while the optimal $\mathbf{K_{PMA}}$ is computed as a function of $\boldsymbol{\Theta}$ by solving an inner control optimization problem \cite{fathy2001coupling}. The new nested optimization problem can be distinguished from Eq. \ref{eq:generic_optim_funct} using the following expression:
	
	\begin{equation}
	\label{eq:nested_optim_funct}
	f^n\big(\boldsymbol{\Theta},\mathbf{K_{PMA}}(\boldsymbol{\Theta}) \big) = w_S f_S(\boldsymbol{\Theta}) + w_C f_C^n\big(\boldsymbol{\Theta},\mathbf{K_{PMA}}(\boldsymbol{\Theta}) \big)
	\end{equation}

where $f_C^n$ represents the inner \textit{nested} control sub-optimization problem.

To perform the outer optimization routine, function only of the mechanical parameter $\boldsymbol{\Theta}$, a particle-swarm optimization (P.S.O.) algorithm was implemented using the routines available in the \matlab Global Optimization Toolbox \cite{GOT_matlab_2021}. Moreover, due to the independent evaluation of the objective function for each particle, its computation can be performed in parallel for each iteration's swarm. The optimization algorithm was therefore run using ISAE Supaero's scientific super-computer PANDO, specificaly on two of its Intel Skylake 6126 2.6 GHz CPUs, each one having 12 cores, with a total RAM of 96 Gb. Distributing the job among these multiple cores allowed for a drastic reduction of computational time. 

\sloppy In practice, at each iteration the P.S.O. algorithm generates a set of $n_S$ particles (defined as swarm), each one with a specific $\boldsymbol{\check{\Theta}}$ value for the design parameter $\mathbf{\Theta}$. It then evaluates the objective function $f^n$ for each particle, $f^n(\check{\boldsymbol{\Theta}},\mathbf{{K}_{PMA}}(\check{\boldsymbol{\Theta}}))$, and determines the particle which yielded to the best value of the objective function $f^n$. Based on the result of the best particle and the ones around it, the method generates a new swarm for the following iteration. The process is looped until a stopping criteria is reached, in this case when the number of iteration gets its maximum value $n_{iter}$.  A swarm size of $n_S = 24$ particles was selected for the P.S.O., distributing the computation in the same number of PANDO's cores. The maximum number of iterations was set to $n_{iter} = 20$.

While this algorithm handles the outer minimization routine, single non-smooth control optimizations are carried out for each swarm's particle at each iteration. This process aims at evaluating the nested control sub-problem objective function $f_C^n\big(\boldsymbol{\Theta},\mathbf{K_{PMA}}(\boldsymbol{\Theta}))$ as function of the mechanical parameter $\check{\boldsymbol{\Theta}}$ of the particle.

	\begin{figure}[h!]
	\centering
	\includegraphics[width =.8\linewidth]{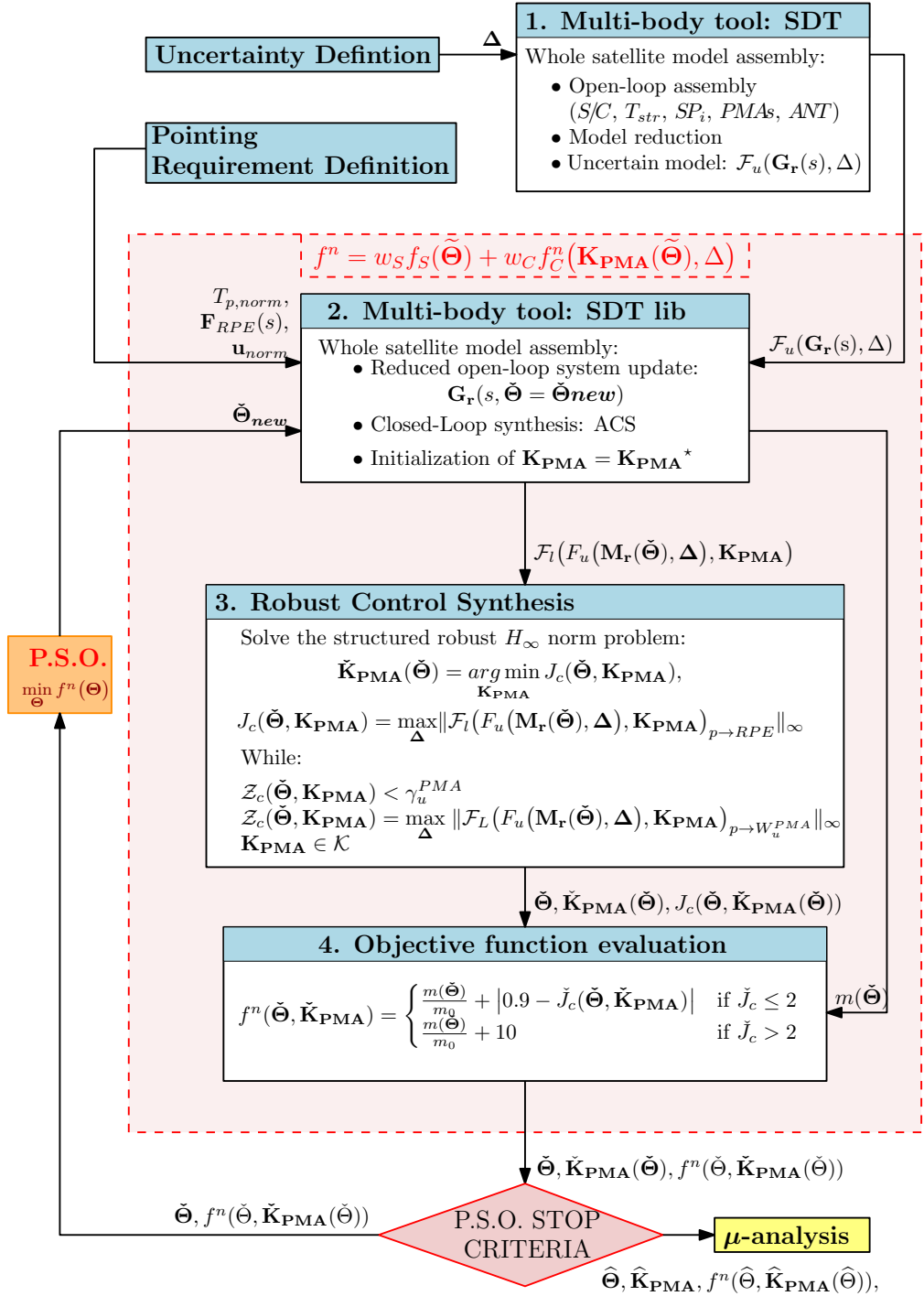}
	\captionof{figure}{\label{fig:Optimization} Co-Design optimization block diagram}
\end{figure} 

Each swarm particle has a single value of the mechanical parameter $\boldsymbol{\check{\Theta}}$ generated by the P.S.O. The process with which the objective function $f^n$ is evaluated for each one of these particle is shown in Fig. \ref{fig:Optimization} and can be summarized in a few key steps:

	\begin{itemize}
		\item  \textit{Step 1. Open-Loop System}. Using the process described in Sect. \ref{sec:OL_sys} for modeling the satellite, the reduced order open-loop system $\mathcal{F}_u \big(\mathcal{G}_r(\mathbf{\Theta}), \boldsymbol{\Delta} \big)$ is obtained.
		
		\item \textit{Step 2. Closed-Loop System}. The closed-loop system of Sect. \ref{sec:CL_sys} is assembled for each particle as follows:
		\begin{enumerate}
			\item The $\boldsymbol{\check{\Theta}}_{new}$ value imposed by the P.S.O. updates the open-loop system 
			$\mathcal{F}_u \big(\mathcal{G}_r(\boldsymbol{\check{\Theta}} = \boldsymbol{\check{\Theta}}_{new}), \boldsymbol{\Delta} \big)$.
			\item The closed-loop is assembled using the new open-loop system, the performance requirements and a custom ACS controller synthesized with the new global inertia matrix.
			\item At the first iteration, all controllers are initialized with all gains at -1. For all the subsequent iterations $i$ of the P.S.O., the controllers are initially set as the controller associated with the best particle found at iteration $i-1$: $\mathbf{K_{PMA}}^\star = \mathbf{\check{K}_{PMA}}^{i-1}$.
		\end{enumerate}
		
		\item \textit{Step 3. Robust Control Synthesis}. The non-smooth parametric robust structured $H_\infty$ problem \cite{7027164} is solved using \matlab's  \verb|systune| routine for each particle having $\boldsymbol{\check{\Theta}}$ to perform the nested optimization for the control sub-problem. This effectively computes the sub-optimal controller $\mathbf{\check{K}_{PMA}}$ which verifies:
		\begin{equation*}
		\mathbf{\check{K}_{PMA}}(\boldsymbol{\check{\Theta}}) = \underset{\mathbf{K_{PMA}}}{\text{arg} \, \text{min}}  \, {J}_c(\boldsymbol{\check{\Theta}}, \mathbf{K_{PMA}}) \; s.t. \;
		\begin{cases}
		{\mathcal{Z}}_c < \gamma_{u}^{PMA} & \text{(Hard Constraint)}\\
		\mathbf{K_{PMAs}} \in \mathcal{K} \\
		\end{cases}
		\end{equation*}
		\sloppy where the perturbation rejection performance on the relative pointing error is given in terms of the previously defined worst-case $H_\infty$-norm
		${J}_c (\boldsymbol{\check{\Theta}}, \mathbf{{{K}_{PMA}}})$, while the effort constraint is given by ${\mathcal{Z}}_c (\boldsymbol{\check{\Theta}}, \mathbf{{{K}_{PMA}}})$. 
		Finally, the weight $\gamma_u^{PMA} = 1$ was selected.
		
		\item \textit{Step 4. Objective function evaluation}. The objective function contains the result of the nested optimization ${J}_c(\boldsymbol{\check{\Theta}}, \mathbf{\check{K}_{PMA}})$ and the mass of the satellite when $h = \boldsymbol{\check{\Theta}}$, computed by inverting the DC gain between $\mathbf{W_{P_1}}$ and $\mathbf{\ddot{u}_{P_1}}$ in the closed-loop model.
		The objective function chosen for this study is again composed by the sum of a structural sub-problem $f_S(\mathbf{\check{\Theta}})$ and a nested control sub-problem $f^n_C (\boldsymbol{\check{\Theta}}, \mathbf{\check{K}_{PMA}})$:
		\begin{equation}
		f^n(\boldsymbol{\check{\Theta}}, \mathbf{\check{K}_{PMA}}) = 
		\begin{cases}
		\frac{m(\boldsymbol{\check{\Theta}})}{m_0} + \big|\, 0.9- {J}_c(\boldsymbol{\check{\Theta}}, \mathbf{\check{K}_{PMA}}) \, \big| & \text{if ${J}_c(\boldsymbol{\check{\Theta}}, \mathbf{\check{K}_{PMA}}) \le 2$}\\
		\frac{m(\boldsymbol{\check{\Theta}})}{m_0}   + 10  &\text{if ${J}_c(\boldsymbol{\check{\Theta}}, \mathbf{\check{K}_{PMA}}) > 2$}\\
		\end{cases} 
		\end{equation}
	
		where $m_0$ is the mass of the system associated with $\boldsymbol{\Theta}_{max} = \text{sup}\ \mathcal{H} = 3\, cm$ in the baseline design to be optimized.
	
			\end{itemize}
		

		We can note that the structural sub-problem, given by $f_S(\mathbf{\check{\Theta}}) = m(\check{\boldsymbol{\Theta}})/ m_0$, aims at reducing the normalized mass of the truss structure. The control sub-problem $f^n_C (\boldsymbol{\check{\Theta}}, \mathbf{\check{K}_{PMA}})$, on the other hand,  aims to stabilize the worst-case performance at 90\% of the pointing requirement, to account for safety margins. Moreover, it penalizes all the solutions with an excessively high values of ${J}_c$, to avoid solutions with extremely low mass and pointing performances way above requirements.
	
	The function $f^n$ is evaluated for each particle of the swarm and feedbacked to the particle swarm algorithm until the maximum number of iterations is reached. The optimized co-design is obtained in the form of the optimal mechanical parameter $\boldsymbol{\widehat{\Theta}}$ and the optimal gains of the PMA controller $\mathbf{\widehat{K}_{PMA}}$. 

\subsection{Co-design optimization results}
\label{sec:results}
\begin{figure}[h!]
	\centering
	\begin{subfigure}{.495\textwidth}
		\includegraphics[width =\textwidth,trim=20 0 20 20,clip]{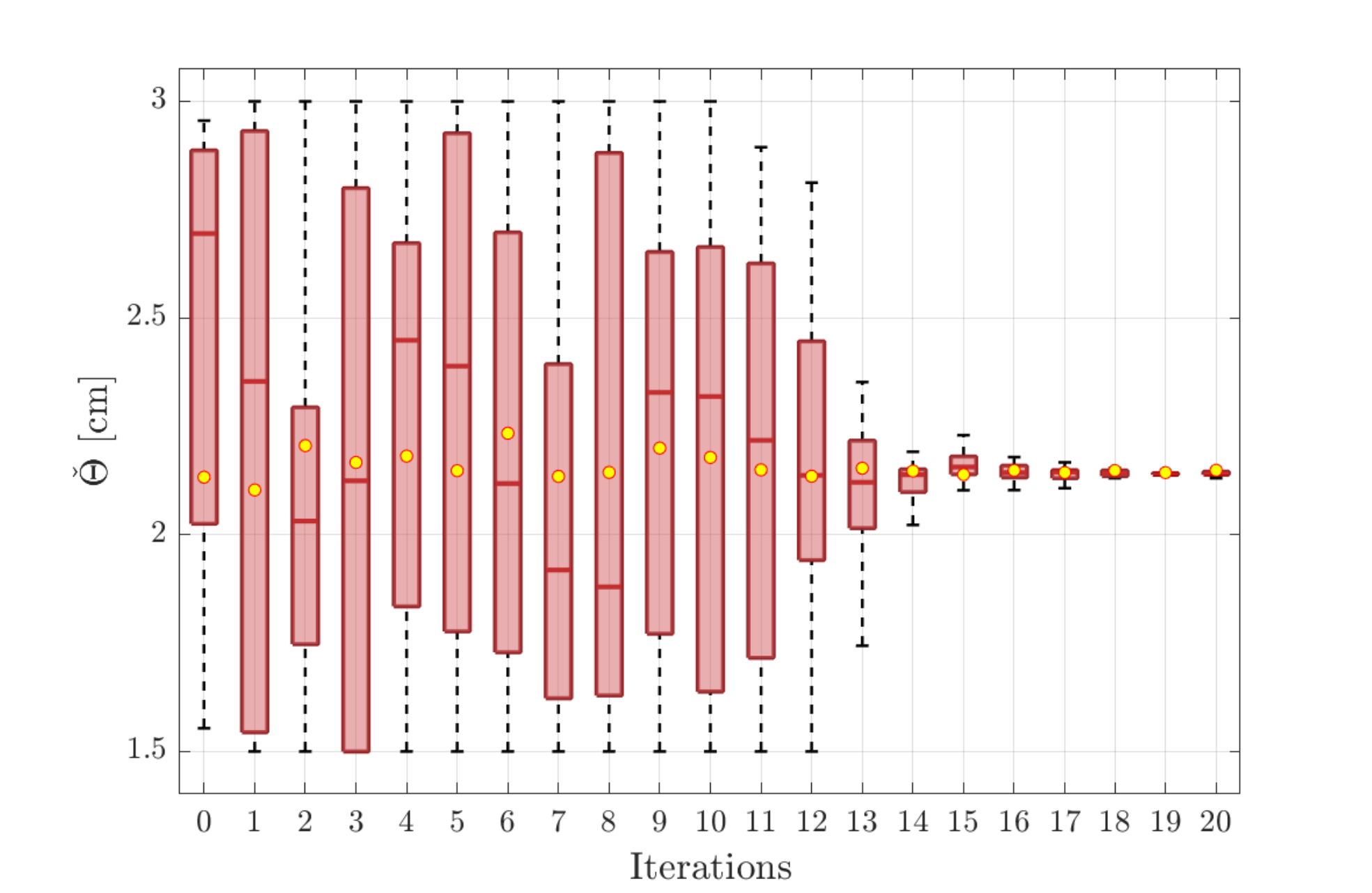}
		\caption{}
		\label{fig:PS_1}
	\end{subfigure}
	\begin{subfigure}{.495\textwidth}
		\includegraphics[width =\textwidth,trim=20 0 20 20,clip]{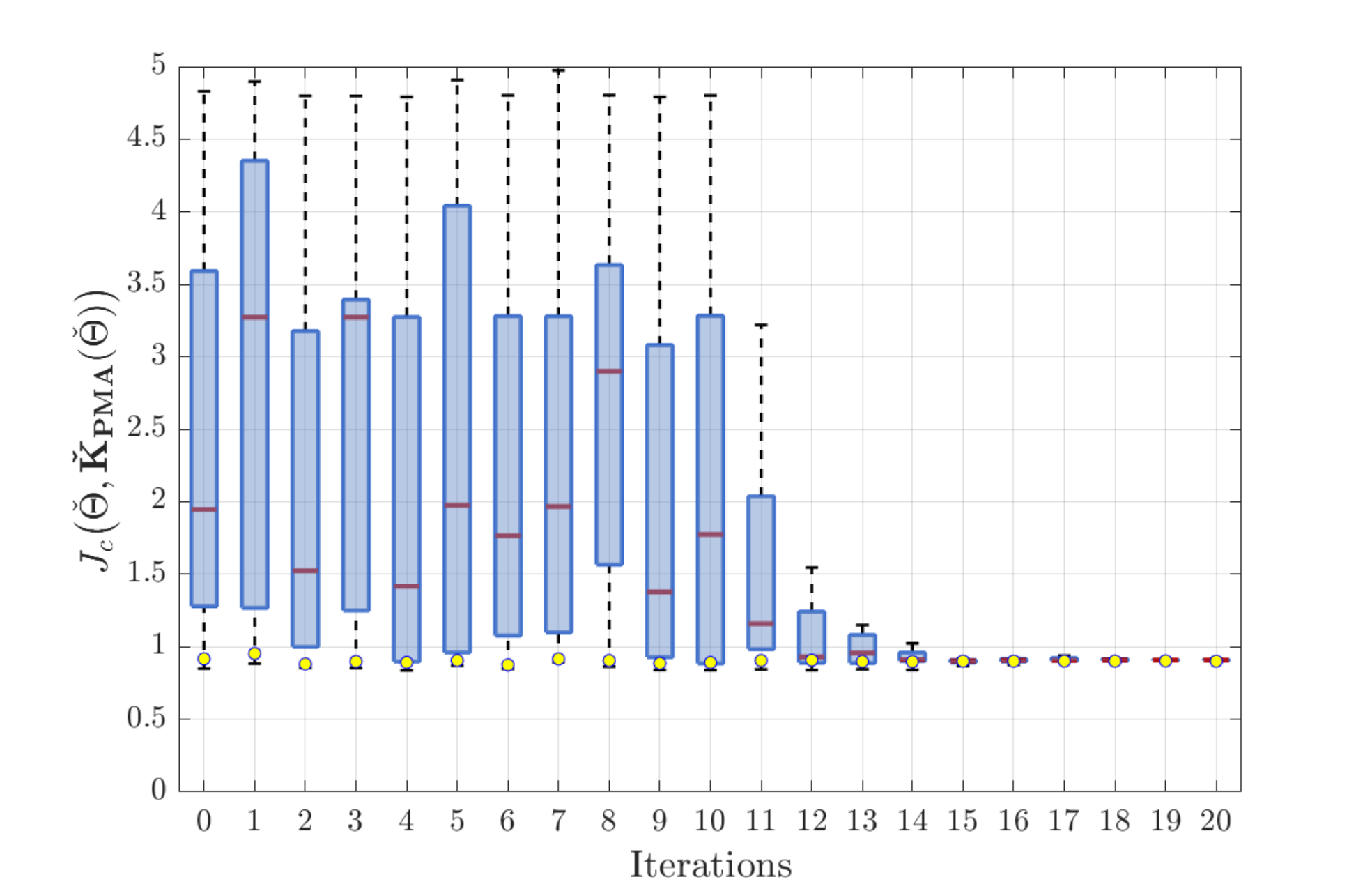}
		\caption{}
		\label{fig:PS_2}
	\end{subfigure} \\
	\begin{subfigure}{.495\textwidth}
		\includegraphics[width =\textwidth,trim=20 0 20 20,clip]{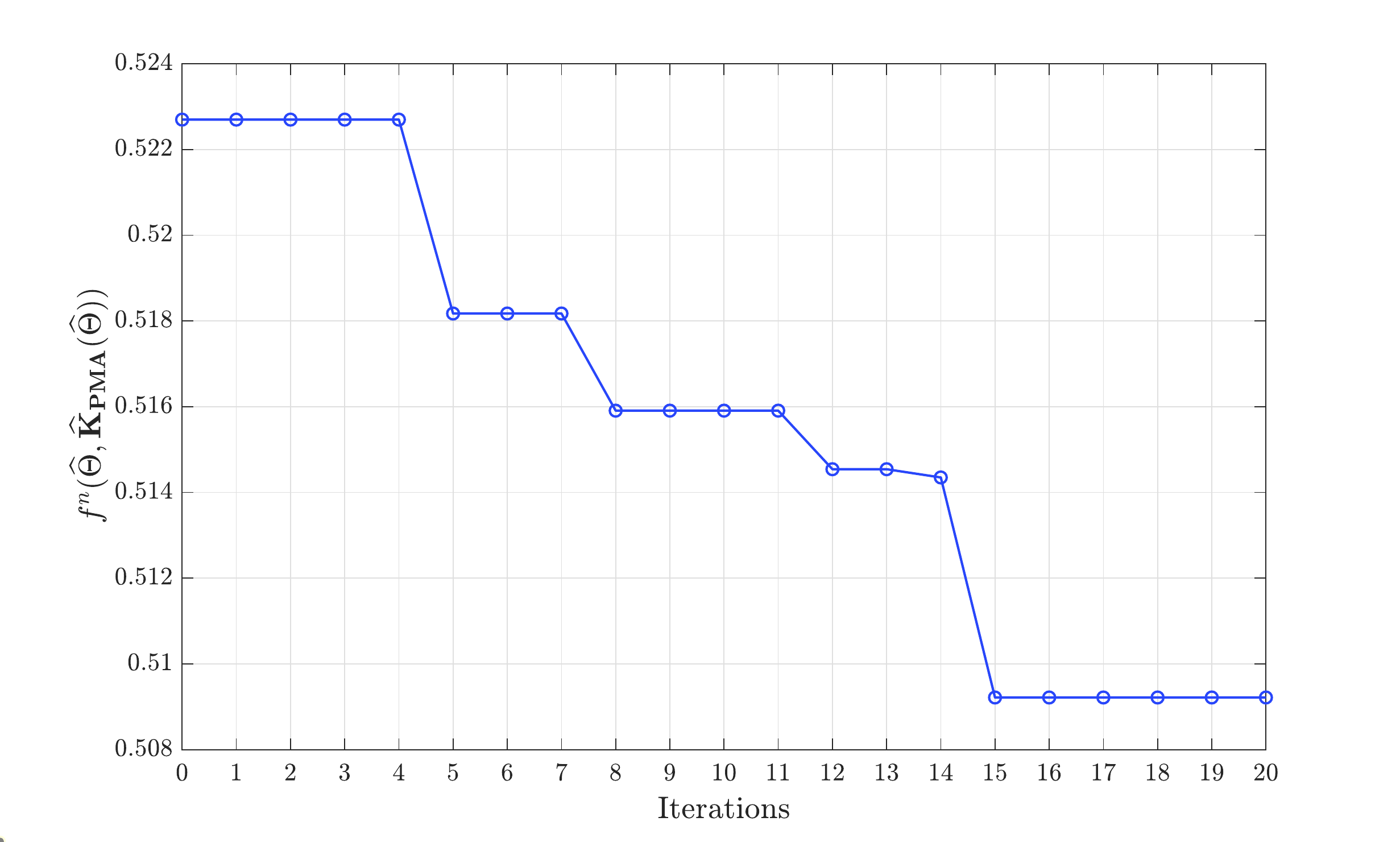}
		\caption{}
		\label{fig:PS_3}
	\end{subfigure}
	\begin{subfigure}{.495\textwidth}
		\includegraphics[width =\textwidth,trim=20 0 0 20,clip]{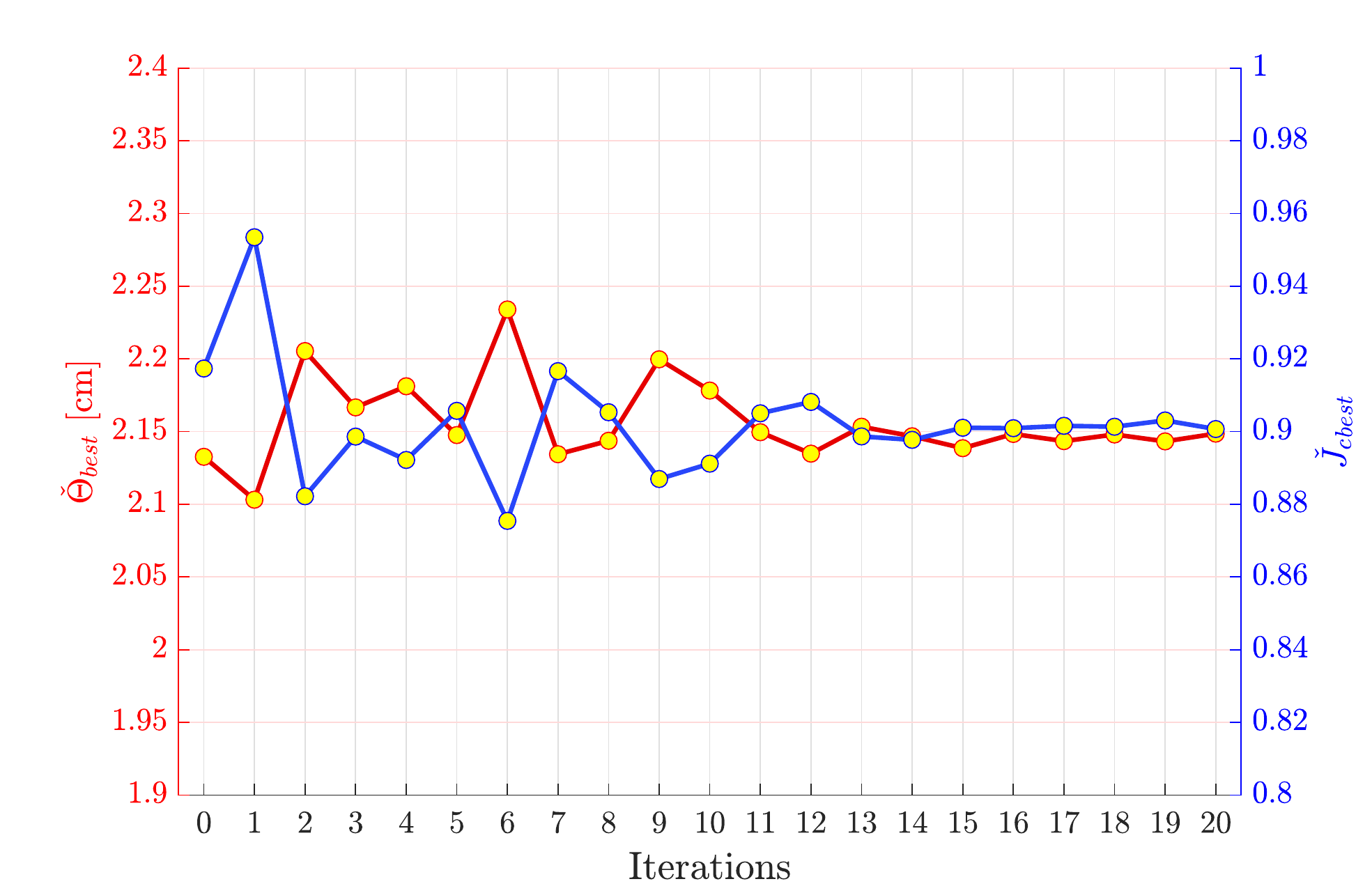}
		\caption{}
		\label{fig:PS_4}
	\end{subfigure} \\
	\caption{Particle swarm optimization results. (a) Particle distribution in terms of the mechanical parameter $\boldsymbol{\check{\Theta}}$ at each iteration. (b) Particle distribution in terms of the optimized control performance ${{J}_c}(\boldsymbol{\check{\Theta}}, \mathbf{\check{K}_{PMA}}) $ at each iteration. (c) Global optimization convergence by means of the objective function $f^n$. (d) $\boldsymbol{\check{\Theta}} $ and ${{J}_c}(\boldsymbol{\check{\Theta}}, \mathbf{\check{K}_{PMA}}) $ for each \textbf{best particle} in terms of $f^n$ at each iteration. Note that the yellow points in (a), (b) and (d) represent the best particles per iteration.}
	\label{fig:ParticleSwarm}
\end{figure}

The results of the co-design optimization process are hereby presented. The four plots in Fig. \ref{fig:ParticleSwarm} showcase the results of the P.S.O. algorithm. As it can be remarked in these figures, the particle swarm algorithm successfully converged towards a global optimal co-design result. This yielded to a mass reduction of 48.91 \% of the T-Truss structure original mass, saving a total of 89.56 kg with respect to the initial design having $m_0 = 183.11$ kg. The full characterization of the optimal result can be found in Table \ref{tab:optim_results}.

\begin{table} [t!]
	\caption{Particle swarm optimization results in terms of $\boldsymbol{\widehat{\Theta}}$, structural mass reduction of $T_{str}$ and optimal pointing performances $\widehat{J}_c$}	
	\label{tab:optim_results}
	\centering
	\begin{tabular}{c c c c c}
		\toprule
		$\boldsymbol{\widehat{\Theta}} \ [cm]$ & $m(\boldsymbol{\widehat{\Theta}}) \ [kg]$ & $m_0 \ [kg]$ &$\Big( m(\boldsymbol{\widehat{\Theta}}) - m_0\Big)/m_0$ & $\widehat{J}_c$ \\ 
		\midrule
		2.139 & 93.55 & 183.11 & -48.91\% & 0.9011 \\ \bottomrule
	\end{tabular}
\end{table}

The first two plots show the distribution of the particles generated by the P.S.O. in terms of the mechanical parameters $\boldsymbol{\check{\Theta}}$ (Fig. \ref{fig:PS_1}) and the worst-case pointing performance ${J_c}(\boldsymbol{\check{\Theta}}, \mathbf{\check{K}_{PMA}})$, result of the nested optimization problems (Fig. \ref{fig:PS_2}). The statistical data is expressed by means of boxplots, indicating the minimum, median, first and third quartiles and maximum of the data. Moreover, the particles which have led to the lowest evaluation of the objective function $f^n$ have been highlighted as yellow dots at each iteration.

Fig. \ref{fig:PS_1} shows how the algorithm properly sampled the totality of the search domain $\mathcal{H}$ and after around 10 iterations it started to converge towards the optimal solution. The first iterations show a great dispersion of the particles, caused by the unpredictable results of the nested optimization of the control sub-problem. The variability seen in Fig. \ref{fig:PS_2} is actually even more pronounced, as particles with failed $H_\infty$-synthesis and extremely high ${{J}_c}(\boldsymbol{\check{\Theta}}, \mathbf{\check{K}_{PMA}}) $ have been removed from the plots. Despite these difficulties, the algorithm successfully converged towards the optimal solution. This can be appreciated  by analyzing the global optimization trend for $f^n$ found in Fig. \ref{fig:PS_3}, which decreases as the iterations progress. It must be remarked how the decrease in $f^n$ is quite small. This is due to the fact that at each iteration at least one of the 24 particles was found to be close to the final optimal solution.

Despite this proximity, these are solutions that have sub-optimal relative importance given to the structural or control sub-problems. In particular, Fig. \ref{fig:PS_4} shows how $\boldsymbol{\check{\Theta}}$ and ${{J}_c}(\boldsymbol{\check{\Theta}}, \mathbf{\check{K}_{PMA}}) $ evolve at each iteration for the best particle at each iteration. These two are the main variables involved in the evaluation of $f^n$ and they represent the relative importance given to $f_S$ and $f^n_C$. This graph displays the concurrent nature of this co-design problem, as higher values of the mechanical parameter implicate better pointing performances and vice-versa. Moreover, it gives context to the best sub-optimal solutions found at each iteration: multiple sub-optimal solution, even close to the optimal one, present differences in $f^n$ composition with respect to $f_S$ and $f^n_C$ which are not optimal.


In terms of computational efficiency, the overall optimization has totally taken $t_{tot} = 11\ h \ 16 \ min$ to converge to the final result, with an average of around $30 \min$ for each iteration. This time is taken almost entirely by the evaluation of the objective function and the resolution of its nested $H_\infty$ synthesis. This global time windows, comparable to the ones of direct co-designs based on \cite{gahinet2011structured}, is possible only thanks to the parallel tasks distribution performed on PANDO's supercomputer. This has in fact enabled the drastic reduction of the total computational duration of at least a 24 factor, as each iteration would have needed to evaluate each particle's $f^n_C$ sequentially.

\subsubsection{Worst-case pointing analysis}\label{sec:mu}




Since the model used to synthesize the robust optimal controller $\mathbf{\widehat{K}_{PMA}} $ is $\mathcal{G}_r(\mathbf{\Theta}) = \mathcal{F}_u (\mathbf{G_r}(\mathrm{s}), \mathbf{\Theta})$, a last validation consists in proving the robustness of the solution when the open-loop system taken into account is built from $\mathbf{G_r}_{low}(\mathrm{s})$.  The latter represents the reduced order system where only the numerical errors at low frequency are truncated while all the high frequency content is kept. Moreover since robust analysis algorithms based on $\mu$-analysis suffer in case of highly repeated parametric uncertainties, a family of $N_\tau$ closed-loop models $P_i(\tau_i, \boldsymbol{\Delta}_p)$ are considered varying the $\tau$ parameter.  In particular, each model $P_i(\tau_i, \boldsymbol{\Delta}_p)$ corresponds to the original system evaluated for $\tau = \tau_i$ (with $\tau_i \in [0,1],\, i=1,...,N_\tau$):
\begin{equation}
\mathcal{P}_i(\tau_i, \boldsymbol{\Delta}_p) = \mathcal{F}_l\left( \mathcal{F}_u \left(\mathcal{G}_{r_{low}}(\widehat{\boldsymbol{\Theta}},\tau_i),\boldsymbol{\Delta}_p
\right),{\mathbf{\widehat K_{PMA}}}
\right)
\end{equation} 

The worst-case gain analysis has been carried out for both transfers $\widetilde{\mathbf{p}} \rightarrow \mathbf{RPE}$ and $\widetilde{\mathbf{p}} \rightarrow \widetilde{\mathbf{W}}_u$. This corresponds to computing the structured singular value upper bound for the $N_\tau$ models:

\begin{equation}
	\text{sup}_{\boldsymbol{\Delta}_p} \mu_{\boldsymbol{\Delta}} = \underset{\boldsymbol{\Delta}_p}{\text{sup}} \ \Big\Vert \mathcal{P}_i(\tau_i, \boldsymbol{\Delta}_p)_{\widetilde{\mathbf{p}} \rightarrow \mathbf{X} } \Big\Vert_\infty 
\end{equation}

and the lower bound:
\begin{equation}
\text{inf}_{\boldsymbol{\Delta}_p} \mu_{\boldsymbol{\Delta}} = \underset{\boldsymbol{\Delta}_p}{\text{inf}} \ \Big\Vert \mathcal{P}_i(\tau_i, \boldsymbol{\Delta}_p)_{\widetilde{\mathbf{p}} \rightarrow \mathbf{X} } \Big\Vert_\infty 
\end{equation}

where the transfer's output $\mathbf{X}$ can be both $\mathbf{RPE}$ or $\widetilde{\mathbf{W}}_u$.

\begin{figure}[h!]
	
	\begin{subfigure}{\textwidth}
		\centering
		\includegraphics[width =\columnwidth]{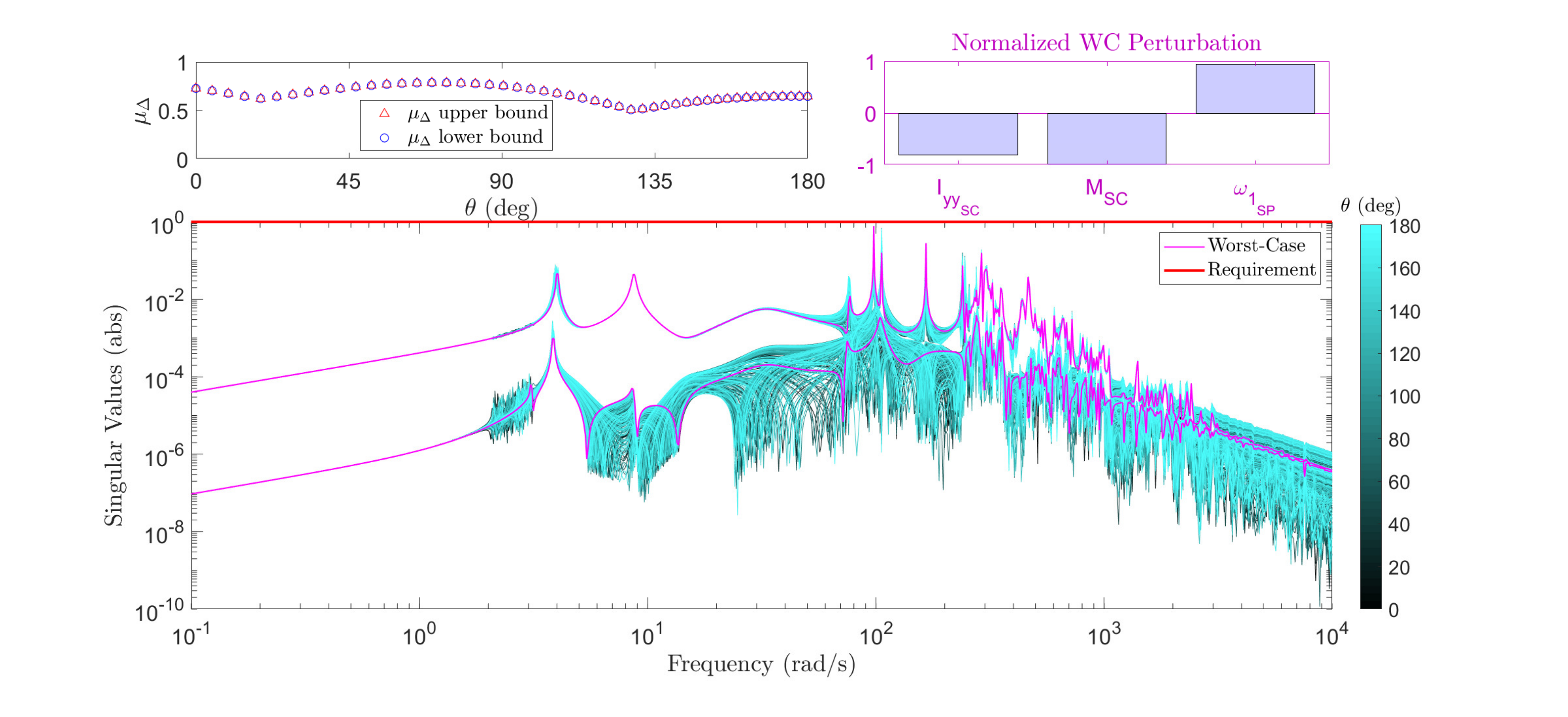}
		\caption{}
		\label{fig:verif_RPE}
	\end{subfigure} \\
	\begin{subfigure}{\textwidth}
		\centering
		\includegraphics[width =\columnwidth]{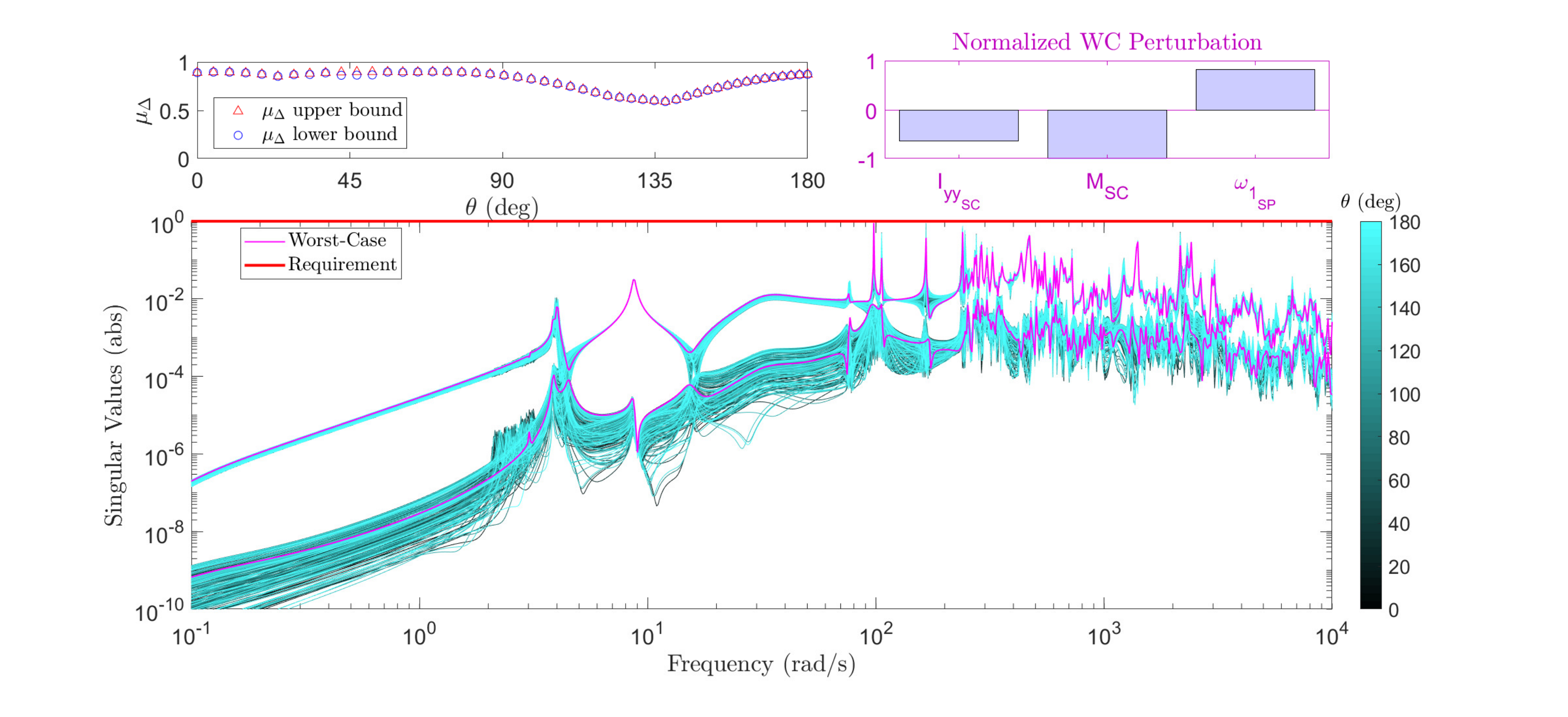}
		\caption{}
		\label{fig:verif_PMAs}
	\end{subfigure}
	\caption{Worst-case analysis for the uncertain optimal solution $\mathcal{P}_i(\tau_i, \boldsymbol{\Delta}_p)$ as a function of $\tau$ for the two transfer functions: (a) $\widetilde{\mathbf{p}} \rightarrow \mathbf{RPE}$ and (b) $\widetilde{\mathbf{p}} \rightarrow \widetilde{\mathbf{W}}_u$. On the top part of the figure, the $\mu$-bounds are displayed together with the worst-case parameter combinations. Down below, the singular values of the system and its worst-case are showcased.}
	\label{fig:mu_analysis}
\end{figure}

The computation of the worst-case bounds is obtained by means of the \verb|wcgain| routine in \matlab's \textit{Robust Control Toolbox} \cite{balas2007robust}.  The upper and lower bounds are showed in Fig. \ref{fig:mu_analysis} for 50 values of $\tau \in [0,1]$. This subset for $\tau$ has been chosen to account for the symmetric configuration of the model in the $\theta \in [0, 180]^\circ$ and $\theta \in [-180, 0]^\circ$ intervals. The worst case $\mu_{\boldsymbol{\Delta}_p}$ bounds are presented in Table \ref{tab:worst_case}.

\begin{table}
	\caption{Worst-case $\mu_{\boldsymbol{\Delta}_p}$ scenarios for the two transfers of interest.}
	\label{tab:worst_case}
	\resizebox{\textwidth}{!}{%
	\begin{tabular}{l l l l l l l l}
	\toprule
	& $\text{sup}_{\boldsymbol{\Delta}_p} \mu_{\boldsymbol{\Delta}_p}$ & $\text{inf}_{\boldsymbol{\Delta}_p} \mu_{\boldsymbol{\Delta}_p}$ & Critical freq. [rad/s] & Critical $\theta$ [deg] & Critical $I_{yy}\ [kg \, m^2]$ & Critical $M_{SC}$ [kg] & Critical $\omega_{1,S\!P}$ [rad/s] \\
	\midrule
	$\widetilde{\mathbf{p}} \rightarrow \mathbf{RPE}$ & 0.7882 & 0.7880 & 97.7440  & 69.4161 &  919.6964 & 720 & 2.9853 \\
	$\widetilde{\mathbf{p}} \rightarrow \mathbf{\widetilde{W}}_u$ & 0.9000 & 0.8993 & 97.7665 & 69.4161 & 937.7710 & 720 &  2.9262 \\ \bottomrule
	\end{tabular}
}
\end{table}

In Fig. \ref{fig:mu_analysis} a complete representation on the worst-case analysis results is presented for the two transfers of interest in terms of their $\mu$-bounds as a function of the angle $\theta = 4 \tan^{-1}(\tau)$, their frequency response and the worst case parameter combinations. It can be seen that the $\mu$-analysis identified the two bounds close to each other and always with a value below the unity, successfully validating the robustness of the design. Moreover, it can be remarked that the bounds show a visible trend, which justifies the choice of a relative low number of $N_\tau$ for this analysis. Furthermore, the study of the singular value responses reveals that the worst-case frequency response of the system remains always below requirement.


\section{Conclusions}
This paper aimed at introducing new analytical tools to model large complex truss structures for space applications in the TITOP/NINOP framework with the specific objective of developing models for structure/control co-design and robust analysis and control. A series of 2D mechanism blocks has been introduced to build a unitary 3D cubic element which serves as a building block for truss structures in a sub-structuring approach. The analysis displayed the potentialities of the approach, as large structures composed by a high number of beam elements can be easily assembled by using blocks of decreasing complexity. Furthermore, a full validation campaign proved the accuracy of these models.

A case study was then introduced to represent the strengths of the TITOP/NINOP approach in performing robust structure/control co-design in presence of parametric uncertainties. A complex 3D truss structure was built using the previously introduced cube elements and attached to a spacecraft to act as support for an high precision antenna. The objective of the co-design was to reduce the structural mass of the system while satisfying a fine pointing requirement. This study case highlighted the potential of these analytical blocks in performing complex multi disciplinary optimization problems. The implementation of a global optimization routine to solve the MDO problem using parallel computation allowed for computational cost reduction and brought to a mass saving of almost 50\% of the original structural mass, while coping with stringent pointing performances and a large set of uncertainties in the mechanical design parameters. 

\begin{appendices}
\begin{table} [h!]

	\caption{Spacecraft mechanical data}
	\label{tab:Sat_prop}	
	\centering
	\resizebox{\textwidth}{!}{
	\begin{tabular}{p{1.5cm} l l r }
		\toprule
		& \textbf{Parameter} & \textbf{Description}  & \textbf{Value \& Uncertainty} \\
		\midrule
		\multirow{5}{1.5cm}{\centering Spacecraft ${S/C}$}
		& $P_1$ & Spacecraft C.o.G. & $[0,\, 0,\, 0]$ m \\
		&  $M_{S\!C}$ & Mass & $800\ kg \pm 20\%$ \\
		& $\begin{bmatrix}
			I_{xx_{S\!C}} & I_{xy_{S\!C}} & I_{xz_{S\!C}} \\
			 &  I_{yy_{S\!C}} & I_{yz_{S\!C}} \\
			 &  &  I_{zz_{S\!C}} \\
		\end{bmatrix}$ & Inertia in $\mathcal{R}_{SC}$ frame 
		& $\begin{bmatrix}
		1000 & 0 & 0 \\
		 &  1000 \pm 20\% & 0 \\
		 &  &  200 \\
		\end{bmatrix}\, kg \, m^2 $ \\ \midrule
		\multirow{10}{1.5cm}{\centering Solar Panels ${S\!P}$}
		& $\mathbf{r}_{OG}^{{S\!P}}$ & Solar panel C.o.G. in $\mathcal{R}_{SP}$ & $[0;-2;0.03] \ m$ \\
		& $M_{{SP}_1}$ & Mass & $80\ kg$ \\
		& $\begin{bmatrix}
		I_{xx_{SP}} & I_{xy_{SP}} & I_{xz_{SP}} \\
		&  I_{yy_{SP}} & I_{yz_{SP}} \\
		&  &  I_{zz_{SP}} \\
		\end{bmatrix}$ & Inertia in $\mathcal{R}_{SP}$  
		& $\begin{bmatrix}
		80 & 0 & -0.1 \\
		&  20  & 22 \\
		&  &  100 \\
		\end{bmatrix}\, kg\ m^2 $ \\
		& $[\omega_{1_{S\!P}},\omega_{2_{S\!P}},\omega_{3_{S\!P}} ]$ & Flexible modes' frequencies & $[2.51 \pm 20\%,\ 3.77, \ 9.42]\ rad/s$ \\ 
		& $[\xi_{1_{S\!P}},\xi_{2_{S\!P}},\xi_{3_{S\!P}} ]$ & Flexible modes' damping & $0.003$ \\ 
		& $\mathbf{L}_{S\!P}$ & Modal participation factors & 
		$\begin{bmatrix}
		-0.002 & -1.5 & -5 & 14 & 0.02 & -0.01 \\
		5	& 1 & -0.1 & 0 & 2 & 15 \\
		0.3 & 0.002 & 0.03 & -0.02 & 3.2 & -0.2 \\
		\end{bmatrix}$
		\\ \midrule
		
		\multirow{4}{1.5cm}{\centering HPP Antenna ${ANT}$}
		& $M_{A\!N\!T}$ & Mass & $20\ kg$ \\
		& $\begin{bmatrix}
		I_{xx_{A\!N\!T}} & I_{xy_{A\!N\!T}} & I_{xz_{A\!N\!T}} \\
		&  I_{yy_{A\!N\!T}} & I_{yz_{A\!N\!T}} \\
		&  &  I_{zz_{A\!N\!T}} \\
		\end{bmatrix}$
		& Inertia in $\mathcal{R_{ANT}}$ 
		& $\begin{bmatrix}
		1.32 & 0 & 0 \\
		&  1.32  & 0 \\
		&  &  2.5 \\ 
		\end{bmatrix}\, kg \, m^2 $ \\ \midrule

		\multirow{10}{1.5cm}{\centering Proof Mass Actuators $PMAs$}
		& $M^{{B}}$ & Casing mass & $0.5\ kg$ \\
		& $\begin{bmatrix}
			I_{xx_{\mathcal{B}}} & I_{xy_{\mathcal{B}}} & I_{xz_{\mathcal{B}}} \\
			&  I_{yy_{\mathcal{B}}} & I_{yz_{\mathcal{B}}} \\
			&  &  I_{zz_{\mathcal{B}}} \\
		\end{bmatrix}$ & Casing Inertia in the PMA frame at $G$
		& $\begin{bmatrix}
		5e-3 & 0 & 0 \\
		&  5e-3  & 0 \\
		&  &  1.6e-4 \\
		\end{bmatrix}\, kg\ m^2 $ \\
		& $\mathbf{r}_{0G}^{PMA}$ & PMA C.o.G. in $\mathcal{R}_{\mathcal{B}}$ & $[0; 0.05; 0] \ m$ \\
		& $\mathbf{r}_{0P}^{PMA}$ & PMA connection point in $\mathcal{R}_{\mathcal{B}}$ & $[0; 0; 0] \ m$ \\
		& $\mathbf{v}$ & Spring-damper direction in $\mathcal{R}_{\mathcal{B}}$ & $[0; 1; 0] \ m$ \\
		& $m_p$ & Proof mass & $0.1 \ kg$ \\
		& $k_p$ & Spring stiffness & $10 \ N/m$ \\
		& $d_p$ & Damper & $1.4 \ Ns/m$ \\ \midrule
	\end{tabular}
}
\end{table}
\begin{table} [h!]
	\caption{Definition of connection points for the spacecraft sub-systems in the spacecraft reference frame $\mathcal{R}_{SC} (P_1,\ x,\ y,\ z )$}
	\label{tab:Connect_points}	
	\centering
	\resizebox{\textwidth}{!}{%
		\begin{tabular}{l l l l || l l l l}
			\toprule
			Description & Point & Coordinates & Unit & Description & Point & Coordinates & Unit \\
			\midrule
			$T_{str}$ connection to $S/C$ at Node 1	& $P_2$ & $[-0.5,\, -0.5,\, 1]$ & 	m & $T_{str}$ connection to $S/C$ at Node 4	& $P_5$ & $[-0.5,\, 0.5,\, 1]$ & m \\
			$T_{str}$ connection to $S/C$ at Node 2	& $P_3$ & $[ 0.5,\, -0.5,\, 1]$ & 	m & Solar Panel 1 $S\!P_1$ connection to $S/C$ & $P_6$ & $[1,\, 0,\, 0]$ & m \\
			$T_{str}$ connection to $S/C$ at Node 3	& $P_4$ & $[0.5,\, 0.5,\, 1]$ & 	m & Solar Panel 2 $S\!P_2$ connection to $S/C$ & $P_7$ & $[-1,\, 0,\, 0]$ & m \\ 
			\bottomrule
		\end{tabular}
	}
\end{table}
\end{appendices}


\bibliography{MyBibliography}

\end{document}